\documentclass[aps,prd,showpacs,twocolumn,amssymb,eqsecnum,nofootinbib,floatfix]{revtex4}

\usepackage{bm}
\usepackage{amsmath}
\usepackage{euscript}
\usepackage{graphicx}

\usepackage[usenames]{color}

\DeclareMathOperator{\sgn}{sgn}

\begin{document}

\title{Verifying black hole orbits with gravitational spectroscopy}

\author{Steve Drasco}
\affiliation{Jet Propulsion Laboratory, California Institute of
Technology, Pasadena, CA 91109}

\date{\today}

\begin{abstract}
Gravitational waves from test masses bound to geodesic orbits of rotating
black holes are simulated, using Teukolsky's black hole perturbation formalism, 
for about ten thousand generic orbital configurations.
Each binary radiates power exclusively in modes with frequencies that are 
integer-linear-combinations of the orbit's three fundamental frequencies. 
General spectral properties are found with a survey of orbits about 
a black hole taken to be rotating at 80\% of the maximal spin.  The orbital eccentricity is varied 
from 0.1 to 0.9.  Inclination ranges from $20^\circ$ to $160^\circ$, and comes 
to within $20^\circ$ of polar.  Semilatus rectum is varied from 1.2 to 3 times the 
value at the innermost stable circular orbits.  The following general spectral properties are found: 
(i) 99\% of the radiated power is typically carried by a few hundred modes, and at most by about a thousand modes, 
(ii) the dominant frequencies can be grouped into a small number of families defined by 
fixing two of the three integer frequency multipliers, and (iii) the specifics of these trends 
can be qualitatively inferred from the geometry of the orbit under consideration. 
Detections using triperiodic analytic templates modeled on these general properties 
would constitute a verification of radiation from an adiabatic sequence of black hole orbits 
and would recover the evolution of the fundamental orbital frequencies.  
In an analogy with ordinary spectroscopy, this would compare to observing the Bohr model's atomic 
hydrogen spectrum without being able to rule out alternative atomic theories or nuclei. 
The suitability of such a detection technique is demonstrated using snapshots computed at 
12-hour intervals throughout the last three years before merger of a kludged inspiral. 
The system chosen is typical of those thought to occur in galactic nuclei, and to be 
observable with space-based gravitational wave detectors like LISA. 
Because of circularization, the number of excited modes decreases as the binary evolves.  
A hypothetical detection algorithm that tracks mode families dominating
the first 12 hours of the inspiral would capture 98\% of the total power over the remaining 
three years.    
\end{abstract}


\maketitle

\section{Introduction} \label{Introduction}

The birth of physics in its modern form can arguably be placed at the
first successful efforts to monitor the motion of the planets as 
they orbit the sun, and to model their planar elliptic orbits.  
Since then, gravitation has transitioned from the best to the least 
tested among the fundamental laws of physics.  Gravitational spectroscopy, 
monitoring power spectra of the outputs from gravitational wave detectors, can  
verify the existence of two-body systems with a dramatically different character.  
Athough characterized by mass ratios similar to those for planetary systems, 
these systems are so distorted by the strong gravity near black hole event horizons that their orbits 
look more like balls of twine than planar ellipses.  These orbits differ significantly from 
those with relativistic precessions which have already been observed and which may 
soon be measured in our galactic center \cite{will 2007}.  Given the relative 
paucity of gravitational experiments, and the enigmatic status of gravitation compared to 
the other fundamental interactions, opportunities to observe these systems are 
highly valuable.  Though unlikely candidates for Brahe-Kepler-like catalysts of another 
revolution in physics, such observations would help to 
elevate tests of gravity to a level that better compares with those for 
the other fundamental physical interactions.   

General relativity predicts that black holes will radiate gravitational 
waves characterized by discrete frequency spectra in at least two broad 
classes of realistically observable scenarios: black hole 
ringdowns, and during extreme mass ratio inspirals (EMRIs).  In both, the 
radiation is a consequence of a single black hole having been slightly perturbed.
Observations of this radiation are therefore ideally suited for studying 
phenomena governed completely by the physics of isolated black holes.

In the case of black hole ringdown, the source of the black hole's perturbation 
is an arbitrary transient event, and the radiation is emitted as the hole 
settles back into a quiescent state.   Ringdown radiation consists of a 
superposition of exponentially damped monochromatic waves, with frequencies and 
damping timescales that are determined by a pair of continuous 
parameters, the black hole's mass and spin.  Potentially observable ringdown 
events include binary systems that merge to form single black holes, and 
supernovae that result in black hole formation.  Ringdown phenomena have been 
studied in detail elsewhere 
\cite{flanagan hughes 1998, dreyer et al, berti et al 2006, berti et al 2007 ringdown}, 
and are not the focus of this work.   

This work instead focuses on EMRIs---black hole binary systems in which the 
mass of the black hole is much greater than the mass of the companion.  
The planned space-based interferometer LISA is anticipated to observe anywhere 
from tens to thousands of EMRIs produced by the capture of compact 
stellar mass black holes (and the occasional white dwarf) by the megamassive 
black holes found in galactic nuclei (masses ranging from $10^5~M_\odot$ to $10^7~M_\odot$) 
with sufficient sensitivity to determine the mass, spin, and quadrupole moment 
of the larger black hole to within a fraction of a percent 
\cite{gair et al 2004, hopman alexander 2006, miller et al 2005, 
sigurdsson 2003, barack cutler 2004, barack cutler 2007}.  
In the recent mock LISA data challenge, a proof of principal 
for a variety of detection algorithms using simulated signals and noise, 
three groups using independent detection algorithms recovered 
EMRI masses and spins to within a few to a few tenths of a percent 
\cite{gair mandel wen, gair babak porter barack, mldc}.

Advanced ground-based detectors (with a lower frequency cutoff of about 10 Hz \cite{advanced ligo}) 
might also observe similar capture events by intermediate black holes in 
globular clusters \cite{brown et al 2006}.
Observation of these intermediate mass ratio inspirals (IMRIs) 
are more speculative, might occur with an estimated upper limit of a 
few to a few tens per year \cite{mandel et al 2007}, and due to their
less extreme mass ratios, may require theoretical waveform models 
that are more sophisticated than those based on black 
hole perturbation theory alone.
See Ref.~\cite{amaro-seoane et al} for a review of EMRI science with both
ground-based and space-based detectors.

Unlike radiation from black-hole ringdown, for EMRIs the source of the black 
hole perturbation is persistent.
On timescales that are short compared to the observable lifetime of the binary,
the companion behaves as a test mass orbiting on a geodesic of the background 
spacetime.  Such orbits are in general characterized by three fundamental 
frequencies \cite{schmidt 2002, mino 2003}, and the corresponding radiation is tri periodic, 
or rather, a superposition of modes that oscillate exclusively at frequencies that 
are integer-linear combinations of the orbit's three fundamental 
frequencies \cite{drasco hughes 2004}
\begin{subequations}\label{waveform model}
\begin{align}
h_{+} - i h_{\times} &= 
\sum_{m kn }
			h_{mkn} e^{-i \omega_{mkn}t}~, & \\
\omega_{mkn} &= m \omega_\phi + k \omega_\theta + n \omega_r ~, \label{frequencies}& 
\end{align}
\end{subequations}
where I am using the convention that all unconstrained summation indices range from $-\infty$ to $\infty$.
Here $\omega_\phi$, $\omega_\theta$, and $\omega_r$, are the fundamental orbital 
frequencies associated with motion in the (Boyer-Lindquist) coordinate shown as a 
subscript, $h_+$ and $h_\times$ are the two independent components of the metric 
perturbation measured by a distant observer, and $h_{mkn}$ are complex amplitudes 
that depend on the observer's position, and on the parameters of the binary 
(masses, spin, and orbit geometry). 

The radiation described by Eq.~(\ref{waveform model}) is only a snapshot of a complete
EMRI waveform.  While work toward precision waveforms for non-test-mass motion throughout
the lifetime of EMRIs remains an active field (see the discussion of Capra waveforms in Ref.~\cite{drasco 2006}),
for short enough times these waveform snapshots are as exact as any envisioned.
For example, for the case of a mass $\mu$ on a circular orbit of initial radius $r$ about a nonrotating 
hole with mass $M \gg \mu$, the normalized overlap of the radiation from geodesic motion alone and that from the true inspiraling motion will be greater than 95\% for times shorter than \cite{drasco 2006}
\begin{equation} \label{dephasing time}
t_\text{dephase} \approx (1 \text{ day}) \left(\frac{M}{10^6 M_\odot}\right) \left(\frac{r}{6M}\right)^{11/4} \left(\frac{M/\mu}{10^5}\right)^{1/2} ~,
\end{equation}
whereas the orbital period is
\begin{equation}
t_\text{orbit} = (8 \text{ minutes}) \left(\frac{M}{10^6 M_\odot}\right) \left(\frac{r}{6M}\right)^{3/2}~.
\end{equation}
Understanding the radiation from simple test mass motion may prove a sufficient basis for useful
observations, however crude.  This is the motivating principal behind the work described here.

This paper has two main results.  The first is an observation of trends in 
a survey of numerically simulated EMRI-snapshots, for thousands of different 
generic orbital configurations, using the code described in 
Refs.~\cite{hughes et al 2005} and \cite{drasco hughes 2006}.  The observed 
trends are summarized as follows: 
(i) When the modes are sorted in order of decreasing power, their power 
decreases fast enough to be primarily confined to a relatively small number of 
modes compared to the number of modes computed when using the algorithm introduced in 
Refs.~\cite{hughes et al 2005} and \cite{drasco hughes 2006}. (ii) The dominant 
modes can be grouped into a small number of families defined by fixing two of the integer 
frequency multipliers.  (iii) The specifics of these trends (for 
example, the precise falloff of power as a function of mode index, or the identity 
of the dominant mode families) can be qualitatively inferred from the geometry 
of the orbit under consideration, as illustrated by the following two examples.

Simple orbits that are very nearly circular are expected to be typical of IMRIs 
that could be found with ground-based detectors.  Mandel \emph{et al}.~estimate 
that the most likely IMRI formation mechanism should result in 
an eccentricity $e < 10^{-4}$ by the time the orbital frequency is brought into 
the observable band above 10 Hz for advanced LIGO \cite{advanced ligo}, and that even for the less likely 
formations mechanisms that can yield higher eccentricities, 90\% of the systems 
should have $e < 0.1$ at 10 Hz \cite{mandel et al 2007}.
For these orbits power falls off as a power law with mode index, and 
99\% of the radiated power is confined to a few to $\sim 10$ modes with frequencies 
(\ref{frequencies}) in the following families:
\begin{equation} 
\label{easy modes}
\omega = \left\{ 
\begin{array}{ll}
        m\omega_\phi & \text{ dominant power at } m=2 \\
        \pm m\omega_\phi + \omega_r & \text{ dominant power at } m=2 \\
        \pm m\omega_\phi + \omega_\theta & \text{ dominant power at } m=1 
\end{array}
\right. ~.
\end{equation} 
where the upper (lower) sign of $\pm$ refers to prograde (retrograde) orbits, and 
mode-power decreases exponentially as $m$ is varied away from 
its value for the dominant mode.  Note that for power spectra, there is no observational 
consequence for a change in a frequency's overall sign.  For any given spectrum, I  
always define mode indices ($m,k,n$) in such a way that the frequency is positive: $\omega_{mkn} > 0$.

More eccentric orbits are expected to be typical of the EMRIs 
observed by space-based detectors.  EMRIs that fall into LISA's frequency band are 
thought to be born with such large initial eccentricities 
($10^{-6} \lesssim 1 - e \lesssim 10^{-3}$) that, although radiation circularizes them, 
even at the time of merger as many as half of the systems should have residual eccentricities 
of about $e \gtrsim 0.2$ \cite{barack cutler 2004}.  
Increasing orbital eccentricity dramatically slows the rate of power falloff
and ultimately results in a spectrum dominated, albeit much less so, by 
the following mode families:
\begin{equation} 
\label{hard modes}
\omega = \left\{ 
\begin{array}{l}
	\pm 2\omega_\phi + n\omega_r \\
       \pm \omega_\phi + \omega_\theta + n\omega_r  \\
       \pm 2\omega_\theta + n\omega_r 
\end{array}
\right. ~.
\end{equation}
where again $\pm$ is $+$ for prograde orbits and $-$ for retrograde orbits,
and mode power decreases as $n$ is varied away from its value $n_{\max}$ at which the 
spectra are peaked.  The value of $n_{\max}$ can be very crudely predicted, with an accuracy 
on the order of 10\%, from eccentricity $e$ according to the formula 
\begin{equation} \label{PM peaks}
n_{\max} \approx \exp(1/2) \left( 1-e\right)^{-3/2}~,
\end{equation}
that was fitted \cite{drasco hughes 2006} to the peaks in the spectra 
derived by Peters and Mathews,  who in 1963 used an analytical treatment of  
Newtonian orbits with the quadrupole formula for gravitational radiation \cite{peters mathews}.
For these eccentric orbits, the rate of power falloff as a function of mode index is
such that as many as $\sim 10^2$ to $10^3$ modes are needed to capture 99\% of the power radiated.
 
The second main result of the paper is a proposal for combining the analytical 
waveform model (\ref{waveform model}) with the observed spectral trends 
so as to create a means for verifying fundamental aspects of black hole physics.  
Precision measurements that 
will test the fundamental physics of black holes have long been among the primary 
motivations for gravitational wave observatories \cite{abramovici et al, lisa science}.  
However, whether or not there are any feasible means for implementing one of the 
more grand measurements, measuring the multipolar 
decomposition of spacetime near black hole candidates in such a way as to identify general relativity as the
only valid theory, has been a subject of 
debate \cite{hughes 2006, psaltis et al 2007}.  
In its most pure form, that measurement will require more than the current 
understanding of what exactly is predicted by general relativity and its 
alternatives.  I propose a related, but less ambitious measurement that will 
constitute a minimal verification of one of general relativity's fundamental predictions 
for black hole perturbation:  that black holes perturbed by bound test masses 
radiate according to the above waveform model (\ref{waveform model}) for times 
that are long compared to the spectrum's fundamental periods 
$2\pi / \omega_{\phi, \theta, r}$ but short compared to the inspiral time.  The measurement would also 
recover the evolutionary sequence of those three fundamental frequencies.  
This is a minimal verification in that it only provides a means for confirming general 
relativity's prediction under the assumption that the perturbed spacetime 
geometry is that of a rotating black hole.  
Observations that provided such a verification may ultimately be subjected to 
the more grand tests based on generalizations \cite{li lovelace 2007} of 
Ryan's theorem \cite{ryan}. In the absence of a greater theoretical understanding
of alternatives to general relativity and black holes ,however, the verification alone 
would not disqualify alternatives for either the background 
spacetime or the theory of gravity.  

The remaining sections of this paper are outlined as follows.  In 
Sec.~\ref{EMRI-snapshot spectra}, I review the relevant equations that define 
generic black hole orbits and that describe the radiation they produce.  In 
Sec.~\ref{Sample spectra}, the radiative spectra for three sample configurations 
are studied and general trends are discussed.  
In Sec.~\ref{Survey of many spectra} I simulate spectra from a grid of orbital configurations 
and discuss how the general trends from the previous section vary across the grid.  
Section \ref{Spectra from a kludged inspiral}
describes how the spectral trends evolve throughout the lifetime of an approximated inspiral.  
In Sec.~\ref{Verification of black hole orbits}, 
I outline a practical means for extracting signals characterized by these 
general spectral trends from data collected by gravitational wave observatories. 
Section \ref{Conclusion} summarizes the paper's main points.

\section{EMRI-snapshot spectra}\label{EMRI-snapshot spectra}

Radiation from generic configurations of test masses bound to black holes has 
been studied in previous work \cite{drasco flanagan hughes 2005, drasco hughes 2006}.  
In this section, I review definitions and equations derived there which will be needed 
throughout the remaining sections.  

\subsection{Orbits}

The physical system described by an EMRI waveform snapshot is a 
nonspinning\footnote{For the EMRIs that could be seen with LISA, the mass ratio renders 
the spin of the smaller object negligible (see Appendix C of Ref.~\cite{barack cutler 2004}).
The same may not be true for the IMRIs that could be seen with advanced ground-based 
detectors, but the effect would likely be competing with more significant complications  
due to the less extreme mass ratios for IMRIs.} 
test mass $\mu$ bound to a black hole with mass $M$ and a spin per unit mass of 
magnitude $0 \le a \le M$.  The test mass' orbit is a bound geodesic of the Kerr 
spacetime determined by $M$ and $a$.  In Boyer-Lindquist coordinates 
$(t, r, \theta, \phi)$, the orbit as a function of proper 
time $x^\mu(\tau)$ satisfies the four first order geodesic 
equations derived by Carter \cite{mtw, carter}.  When bound solutions are 
parametrized as functions of Mino time $\lambda$, with 
$d\tau = (r^2 + a^2 \cos^2\theta)d\lambda$, two of 
their coordinates are periodic \cite{mino 2003}
\begin{subequations} \label{mino orbits}
\begin{align}
t(\lambda) &= \Gamma\lambda 
	+ \sum_{kn}
	t_{kn} e^{-i(k \Upsilon_\theta + n \Upsilon_r) \lambda}~,& \\
r(\lambda) &= 
	\sum_{n}
	r_{n} e^{-in \Upsilon_r \lambda}~,& \\
\theta(\lambda) &= 
	\sum_{k} 
	\theta_{k} e^{-ik \Upsilon_\theta \lambda}~,& \\
\phi(\lambda) &= \Upsilon_\phi \lambda 
	+ \sum_{kn}
	\phi_{kn} e^{-i(k \Upsilon_\theta + n \Upsilon_r)\lambda}~,& 
\end{align}
\end{subequations}
where the coefficients in front of the exponentials are constants (with values that cause these seemingly complex sums to be real).
The quantity $\Gamma$ relates the Mino-frequencies $\Upsilon_{r,\theta,\phi}$ to 
coordinate-time frequencies
\begin{equation}
\omega_{r,\theta, \phi} =  \Upsilon_{r, \theta, \phi} / \Gamma ~, 
\end{equation}
that appear in the radiation observed by distant observers (\ref{waveform model}).  The three 
spatial Boyer-Lindquist coordinates of the orbit 
are not periodic functions of coordinate time $t$, however it follows 
from the formalism in Ref.~\cite{drasco hughes 2004} that 
they have simple biperiodic forms 
\begin{subequations} \label{coordinate orbits}
\begin{align}
r(t) &= 
	\sum_{kn}
	{\tilde r_{kn}} e^{-i(k \omega_\theta + n \omega_r)t}~,& \\
\theta(t) &= 
	\sum_{kn} 
	{\tilde \theta_{kn}} e^{-i(k \omega_\theta + n \omega_r)t}~,& \\
\phi(t) &= \omega_\phi t 
	+ \sum_{kn}
	{\tilde \phi_{kn}} e^{-i(k \omega_\theta + n \omega_r)t}~,& 
\end{align}
\end{subequations}
where the expansion coefficients are again constants.  
A derivation of the coefficients ${\tilde r_{kn}}$, 
${\tilde \theta_{kn}}$, and ${\tilde \phi_{kn}}$ in terms of 
the coefficients in the Mino-time series (\ref{mino orbits}) is given in 
Appendix \ref{append}.

The orbital frequencies $\omega_{r,\theta,\phi}$ are uniquely determined by 
specifying the three constants associated with Killing fields, energy $E$, axial 
angular momentum $L$, and Carter's constant $Q$ 
\begin{align}
E &= - \mu u^t ~,& \\
L &=  \mu M u^\phi ~,& \\
Q &= (r^2 + a^2\cos^2\theta)^2 (u^\theta)^2 & \nonumber \\
  &+ L^2 \cot^2 \theta + a^2(\mu^2 - E^2) \cos^2 \theta~,&
\end{align}
where $u^\alpha = dx^\mu/d\tau$ is the orbit's four-velocity.  
They can also be determined by 
specifying the orbit's coordinate boundaries between two radial turning 
points and between two angular turning points that are symmetric about 
the equatorial plane at $\theta = \pi/2$
\begin{align}
r_{\min}         &\leq r \leq         r_{\max}~,& \\
\theta_{\min} &\leq \theta \leq \pi - \theta_{\min}~,&
\end{align}
or by specifying
three geometric constants generalized from Newtonian orbits: 
eccentricity $e$, semilatus rectum $p$, and inclination $\iota$
\begin{align}
\iota  & = \frac{\pi}{2} - \sgn(\omega_\phi) \theta_{\min}~,& \\
\frac{r_{\min}}{M} &= \frac{p}{1 + e}~,& \\
\frac{r_{\max}}{M} &= \frac{p}{1 - e}~,& 
\end{align}
where $\sgn(\omega_\phi)$ is $1$ for prograde orbits and $-1$ for retrograde orbits.
See Appendix A of Refs.~\cite{schmidt 2002} or \cite{drasco hughes 2006} for
explicit formulae relating the geometric orbital constants, 
the formal Killing constants $E$, $L$, and $Q$, the frequencies 
$\omega_{\phi,\theta,r}$, and the Mino frequencies $\Gamma$ and $\Upsilon_{\phi, \theta, r}$.  
Each of the following sets of parameters are uniquely determined by fixing the values
for any one of them
\begin{align}
&(\Upsilon_\phi, \Upsilon_\theta, \Upsilon_r)~, & \\
&(\omega_\phi, \omega_\theta, \omega_r)~,& \\
&(E, L, Q)~, & \\
&(r_{\min}, r_{\max}, \theta_{\min})~, & \\
&(e, p, \iota)~.& 
\end{align}
Following the terminology of the Guelph group 
\cite{pound poisson nickel 2005, pound poisson 2007a, pound poisson 2007b}, each of 
these triples is a complete set of principal orbital elements.  

After fixing the principal orbital elements, the orbit is not completely determined 
until one specifies an initial position, or some equivalent set of parameters, called 
positional orbital elements \cite{pound poisson nickel 2005, pound poisson 2007a, pound poisson 2007b}.  
Here I will use the following orbital elements 
\begin{equation}
(\lambda_t, \lambda_\phi, \lambda_r, \lambda_\theta)~.   
\end{equation}
These are defined such that, after specifying them and the principal orbital elements, any bound 
black hole orbit can be uniquely expressed as follows 
\begin{subequations} \label{general from fiducial}
\begin{align}
t(\lambda) &= \Gamma(\lambda - \lambda_t) 
        + \sum_{k=1}^\infty \hat t_k^\theta 
        \sin[k\Upsilon_\theta(\lambda-\lambda_\theta)] 
	\nonumber & \\
	& + \sum_{n=1}^\infty \hat t_n^r 
        \sin[n\Upsilon_r(\lambda-\lambda_r)] ~,& \\
r(\lambda) &=  
        \sum_{n=0}^\infty
        \hat r_{n}  \cos[n \Upsilon_r (\lambda-\lambda_r)] ~,& \\  
\theta(\lambda) &= 
        \sum_{k=0}^\infty 
        \hat \theta_{k}  \cos[k \Upsilon_\theta (\lambda-\lambda_\theta)] ~,& \\
\phi(\lambda) &= \Upsilon_\phi (\lambda - \lambda_\phi) 
        + \sum_{k=1}^\infty \hat \phi_k^\theta 
        \sin[k\Upsilon_\theta(\lambda-\lambda_\theta)] 
	\nonumber & \\
	& + \sum_{n=1}^\infty \hat \phi_n^r 
        \sin[n\Upsilon_r(\lambda-\lambda_r)] ~,& 
\end{align}
\end{subequations}
were the hatted coefficients depend only on the principal orbital elements and are given by 
integrals over the $\lambda$-derivatives of the coordinates as given by Eqs.~(2.27) and (2.28) in 
Ref.~\cite{drasco hughes 2006}\footnote{Because of a difference in notation, replace 
$\Delta x$ there with $\hat x$ used here, for $x = t, \phi$.}.  
The first two positional elements,  $\lambda_t$ and $\lambda_\phi$ can take on any value.  The second two are defined on
$0 \leq \lambda_{r} < 2\pi/\Upsilon_r$ and $0 \leq \lambda_{\theta} < 2\pi/\Upsilon_\theta$, respectively as the 
smallest positive values of $\lambda$ at which the coordinate shown in their subscript reaches the smaller of its two turning points.  

\subsection{Radiation}

Black holes that are forever perturbed by bound test masses produce
a gravitational wave field whose two orthogonal linearly polarized components,  $h_+$ and $h_\times$, can be expressed as a single complex function made up of a series of modes that oscillate at frequencies that are integer-linear combinations of the orbit's fundamental frequencies \cite{drasco hughes 2004}.  For  
an observer located at $(t, r, \theta, \phi)$, that function is 
\begin{equation}
h_{+} - i h_{\times} = \sum_{mkn} h_{mkn} e^{-i \omega_{mkn}t}~,
\end{equation}
up to corrections of order $\mu^2/M^2$.  The complex mode amplitudes are given by
\begin{equation} \label{mode amplitudes}
h_{mkn} = -2\frac{e^{i (\omega_{mkn}r + m\phi)}}{r\omega^2_{mkn}}
			\sum_{l = 2}^\infty 
            Z_{lmkn}
            S_{lmkn}(\theta) 
\end{equation}
where $S_{lmkn}(\theta)$ and $Z_{lmkn}$ are quantities that are found by applying Teukolsky's black 
hole perturbation formalism \cite{teukolsky}, as described in Sec.~III of Ref.~\cite{drasco hughes 2006}.  The 
real function $S_{lmkn}(\theta)$, that has here been normalized over the unit sphere  
\begin{equation}
2\pi \int_0^\pi d\theta~ [S_{lmkn}(\theta)]^2 \sin\theta = 1~,
\end{equation}
satisfies Teukolsky's angular equation, a homogeneous ordinary differential equation.
The complex numbers $Z_{lmkn}$ are constant coefficients that come from the limiting behavior 
of the physical solutions to Teukolsky's radial equation, an inhomogeneous ordinary differential equation.
At infinity, that equation's general solutions have the following 
form (see Teukolsky's original papers \cite{teukolsky}, or for a particularly good textbook-level 
discussion, Sec.~4.8.1 of \cite{andersson frolov novikov})
\begin{equation}
R_{lmkn}(r\to \infty) = R^{\text{out}}_{lmkn}r^3 e^{i \omega_{mkn} r^*}
                                + R^{\text{in}}_{lmkn} \frac{1}{r} e^{-i\omega_{mkn} r^*}~.
\end{equation}                                
The solution of Teukolsky's master equation fully describes the system's radiation, to first order in the perturbation,  and is a sum over products of the separated functions 
$R_{lmkn}(r)$, $S_{lmkn}(\theta)$, $e^{im\phi}$, and $e^{-i\omega_{mkn}t}$.
Therefore, at radial infinity, the complex coefficients $R^\text{out}_{lmkn}$ determine the amount of outgoing radiation (toward infinity), while the $R^\text{in}_{lmkn}$ determine the amount of in-going radiation (toward the event horizon).  The complex numbers $Z_{lmkn}$ appearing in the mode amplitudes for the gravitational 
wave (\ref{mode amplitudes}) are taken from the solution that obeys a boundary condition with  zero ingoing radiation at infinity
\begin{equation}
R_{lmkn}(r\to\infty) = Z_{lmkn} r^3 e^{i\omega_{mkn}r}~,
\end{equation}
as well as only in-going radiation at the event horizon.  Details on how to calculate these solutions, and on how the 
horizon-boundary condition enters, can be found in Ref.~\cite{drasco hughes 2006}.

The time-averaged rates of change for the principal orbital elements can be given in terms of 
the quantities described above\footnote{The time averages can be simply defined in terms of Mino-time 
integrals of length $2\pi/\Upsilon_{r}$ and $2\pi/\Upsilon_{\theta}$.  See Sec.~3.8 and 9.1 of Ref.~\cite{drasco flanagan hughes 2005} for details.}.  
Each rate of change can be expressed as two fluxes: an outgoing 
flux at radial infinity and an ingoing flux at the event horizon.  Horizon fluxes are usually two or three orders of magnitude smaller than the fluxes at infinity (see Ref.~\cite{poisson sasaki} for an analytic treatment of circular nonspinning binaries, or Tables IV and VII from Ref.~\cite{drasco hughes 2006} for 
numerical examples of generic binaries).  Though they may prove
observable indirectly through their influence on the orbital evolution \cite{li lovelace 2007}, I will not 
discuss them here apart from mentioning that their calculation is 
nearly identical to that of fluxes at infinity. 

The time-averaged power radiated to infinity at frequency $\omega_{mkn}$, as measured by distant observers,
is given by \cite{drasco hughes 2006} 
\begin{equation}      
\left< \frac{dE}{dt} \right>_{mkn} = \frac{1}{4\pi \omega_{mkn}^2} \sum_{l = 2}^\infty 
                \left| Z_{lmkn} \right|^2 ~,
\end{equation}
and the total time-averaged energy flux at infinity is then just a sum of the power radiated 
at all possible frequencies
\begin{equation}
\left< \frac{dE}{dt} \right> = \sum_{mkn} \left< \frac{dE}{dt} \right>_{mkn}~.
\end{equation}
The remainder of this paper studies the distribution of power among the various terms in this sum 
and how that distribution is affected by the configuration of the orbit.  Understanding this 
distribution is akin to understanding the evolution of the other princple orbital elements, since 
they are all somewhat simply related.  The time-averaged flux of the other two formal constants of 
motion can be written as \cite{drasco flanagan hughes 2005, sago et al 2005, drasco hughes 2006, sago et al 2006} 
\begin{eqnarray} \label{fluxes}
\left< \frac{dL}{dt} \right> &=& \sum_{mkn} \left< \frac{dE}{dt} \right>_{mkn}
   \frac{m}{\omega_{mkn}}\\
\left< \frac{dQ}{dt} \right> &=& 2\sum_{mkn} \left< \frac{dE}{dt} \right>_{mkn}
\left( 
	\frac{m}{\omega_{mkn}} L \left<\cot^2 \theta \right> \right. \nonumber \\
	&+&\left.  \frac{k}{\omega_{mkn}} \mu\Upsilon_\theta 
	 -a^2 E \left<\cos^2 \theta \right>  
\right)~,
\end{eqnarray}
where the angled brackets represent a time average.
From these expressions, one can compute the time-averaged evolution for any 
set of principal orbital elements due to radiation at infinity.  

This all but concludes the review of quantities needed here to describe EMRI snapshots.
Before moving on however, an important property about how the various orbital parameters 
influence the radiation should be addressed.  It has been shown that the positional orbital 
elements have a somewhat simple influence on the radiation, by comparison 
to the influence of the principal orbital elements.  Their only influence on the quantities 
discussed above is as a phase factor in the complex numbers $Z_{lmkn}$.  That factor can be written as \cite{drasco flanagan hughes 2005, mino 2007}
\begin{equation} \label{positional phase}
Z_{lmkn}(\lambda_t, \lambda_\phi, \lambda_{r}, \lambda_{\theta}) = e^{i \chi_{mkn}}
Z_{lmkn}(0, 0, 0, 0) ~,
\end{equation}
where $\chi_{mkn}$ is given by
\begin{equation}
\chi_{mkn} = m \Upsilon_\phi (\lambda_\phi - \lambda_t) 
           + k \Upsilon_\theta(\lambda_\theta - \lambda_t) 
           + n \Upsilon_r(\lambda_r - \lambda_t)~.
\end{equation}
The radiative fluxes for the principal orbital elements (\ref{fluxes}) are therefore independent 
of the positional orbital elements, since they depend only on $|Z_{lmkn}|$.  So the positional orbital elements
will not be relevant to the discussion of power spectra and their evolution in the remaining sections.
However, the mode amplitudes of the waveform (\ref{mode amplitudes}) are functions of 
$Z_{lmkn}$, and not just their moduli.  Therefore, knowing the evolution of the principle orbital 
elements alone is insufficient for evolving from one snapshot to the next in an optimal coherent matched filtering 
detection algorithm.  To evolve waveforms coherently, one needs a prescription for changing both the principal and positional orbital elements.  This issue will be revisited when discussing detection algorithms in Sec.~\ref{Verification of black hole orbits}.

Before discussing the results of the simulations, it is perhaps useful to remind readers who are 
more familiar with other simulations of radiating black hole binary systems that for these snapshot spectra, the source of the radiation is ever-present.  There is no initial data from which imperfections could produce junk radiation which dies out over time.  Of course no source of radiation could really persist forever like this, but that is why these spectra are ``snapshots'' of EMRI spectra.  As described in the introduction, the snapshot spectra match
the true spectra from EMRIs only for sufficiently short observation times, shorter than the dephasing time 
(\ref{dephasing time}).

\section{Sample spectra}\label{Sample spectra}

In this section I describe EMRI snapshot spectra from three sample systems and 
identify properties that are useful for understanding spectra from systems with arbitrary 
orbit geometries.  All of the spectra shown in this paper
were simulated using the numerical code that was first described in Ref.~\cite{drasco hughes 2006}.  
When simulating the spectra described in this paper, 
all adjustable parameters of that code were set to the same values 
used when computing the catalog of orbits introduced in Sec.~V of that paper, with one exception.
The one exceptional code parameter is the requested fractional accuracy $\varepsilon_\text{flux}$ 
in the total radiated power $\left<dE/dt\right>$.  When relevant, the value of $\varepsilon_\text{flux}$ used here 
will be stated below.    

\begin{figure*}
\includegraphics[width = .459\textwidth]{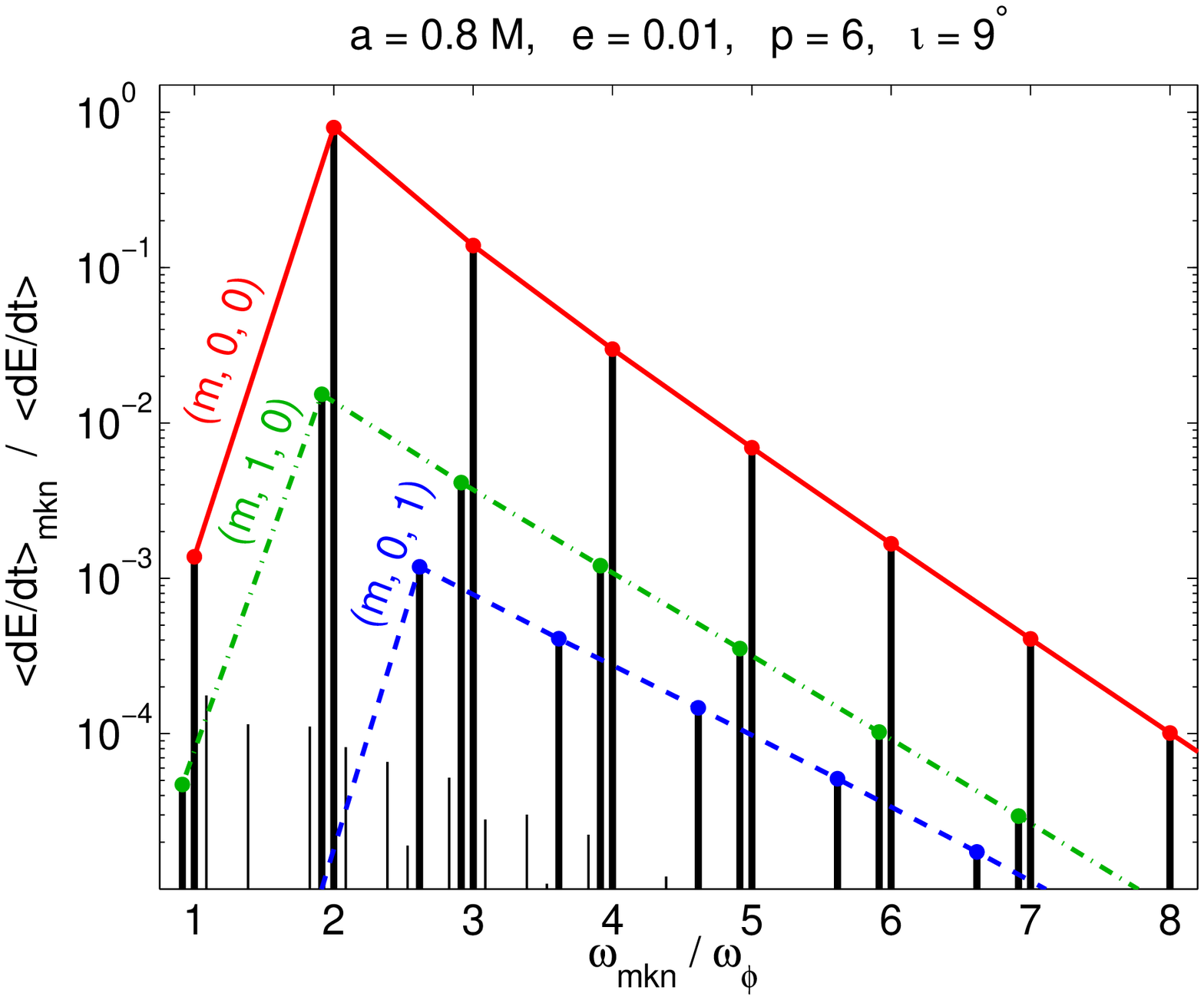}
\includegraphics[width = .459\textwidth]{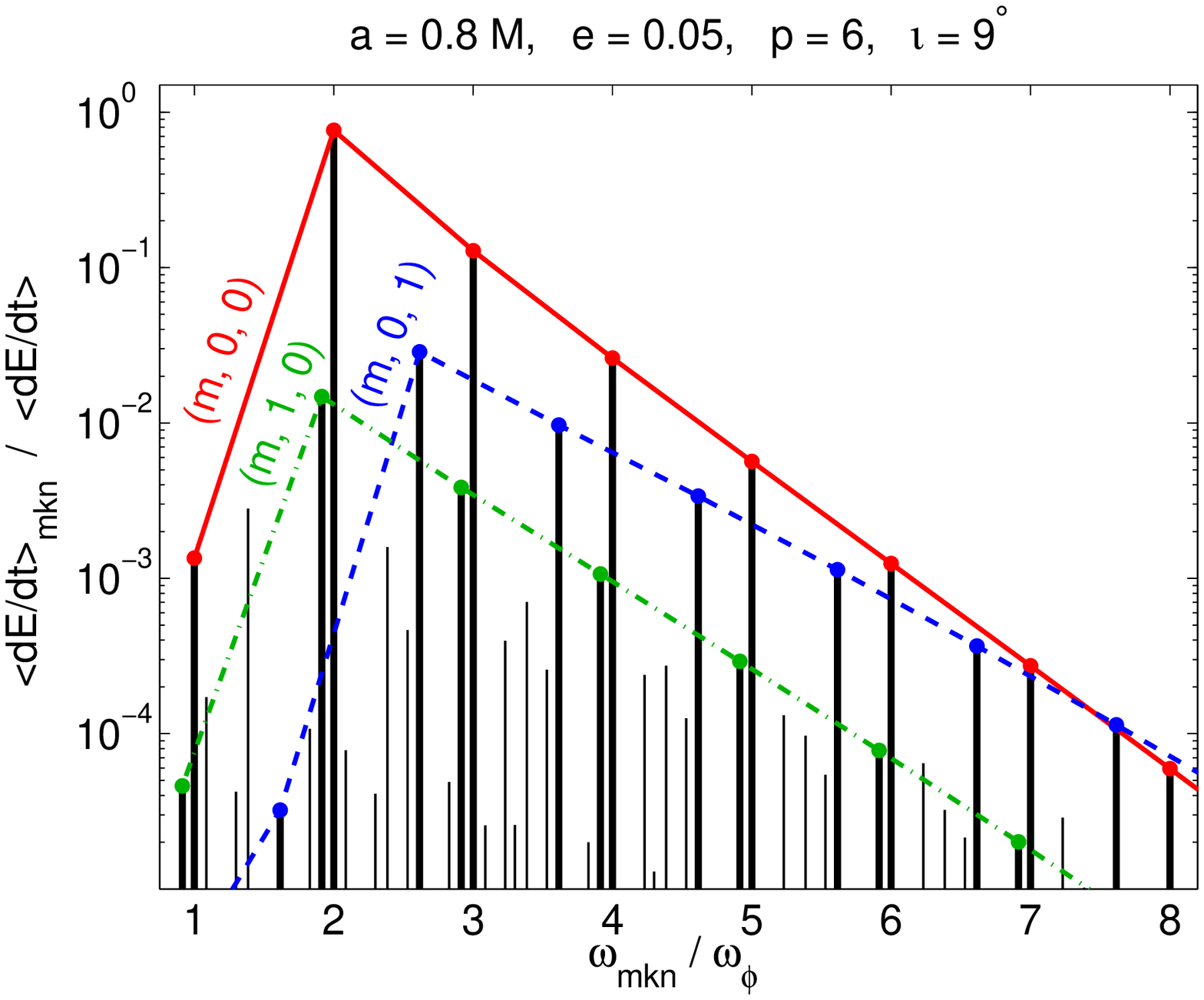}
\caption{\label{simple spectra} The dominant spectral lines for two relatively simple orbital configurations.
The parameters describing the system (black hole spin $a$, eccentricity $e$, semilatus rectum 
$p$, and inclination $\iota$) are shown above each panel.  For the spectrum on the left, 80\% of the total power is 
carried by the dominant mode and 99\% is carried by the most powerful 6 modes.
For the spectrum on the right, 77\% of the total power is 
carried by the dominant mode, and 99\% is carried by the most powerful 11 modes.
For each spectrum, 99.99\%  of the total power is carried by the lines shown. 
The red solid line traces over the tops of the lines carrying radiation at frequencies 
$\omega_{m00}= m\omega_\phi$, for various values of $m$ (which can be read off from the horizontal axis).  
Similarly, the green dash-dot line 
highlights frequencies $\omega_{m10}= m\omega_\phi+\omega_\theta$ (peaked at $m = 1$), and the 
dashed blue line shows $\omega_{m01}= m\omega_\phi+\omega_r$ (peaked at $m=2$).  The thicker spectral 
lines are members of these three mode families, while the thinner lines are not.}
\end{figure*}

Figure~\ref{simple spectra} shows the dominant spectral lines from two relatively simple orbits, both 
computed to a fractional accuracy of $\varepsilon_\text{flux} = 10^{-6}$, in the total 
radiated power $\left<dE/dt\right>$.  The orbits for these two spectra are simple in the sense that 
the motion of the test mass is very nearly restricted to a constant radius, and to the equatorial plane of the large hole.  Correspondingly, these spectra are also somewhat simple.
For both, the peak in the power spectrum occurs at a frequency of $2\omega_\phi$ as one might guess 
from, for example, the waveforms computed by Peters 
and Mathews using Newtonian orbits and the quadrupole formula \cite{peters mathews}.  
The remaining power is distributed predominantly among three families of modes fixing the integer multipliers 
for the radial and azimuthal frequencies to be either zero or one.  The frequencies for those mode families are
\begin{equation} 
\label{simple families}
\omega = \left\{ 
\begin{array}{ll}
        m\omega_\phi & \text{ dominant power at } m=2 \\
        m\omega_\phi + \omega_r & \text{ dominant power at } m=2 \\
        m\omega_\phi + \omega_\theta & \text{ dominant power at } m=1 
\end{array}
\right. ~.
\end{equation}

From Fig.~\ref{simple spectra}, one can see that the power in any given mode falls off exponentially 
with $m$, at a rate that is determined by both the orbit geometry, and the values of $k$ and $n$ that define 
the family.  These mode families turn out to dominate the spectra for all orbits with sufficiently small eccentricity 
and inclination, and the exponential falloff for power in modes within a fixed family turns out to be a general trend
for these simple spectra.

The distribution of power among modes or mode families is determined by the orbital geometry.  
The two panels of Fig.~\ref{simple spectra} show that increasing 
orbital eccentricity draws more power into the family involving the radial frequency, defined by
modes (\ref{frequencies}) with $(m, k, n) = (m, 0, 1)$.
The spectrum from the system with higher orbital eccentricity also has the greater number of excited 
modes which are not members of the three families that dominate simple orbits.  
The following two sections will demonstrate that, in general, the complexity of the spectrum from any EMRI snapshot, 
or the number of modes excited by any fraction 
of the total power, is more sensitively dependent on eccentricity than on inclination or
semilatus rectum.  

This general rule that eccentricity governs spectral complexity is in accordance with the preliminary investigation 
of spectral dependence on orbit geometry given in Ref.~\cite{drasco hughes 2006}.  There, significant 
waveform ``voices'' were defined by sets of frequencies $\omega_{mkn}$ defined as follows
\begin{subequations} \label{voices}
\begin{align}
&\text{azimuthal voice: }& k = 0 \text{ and } n = 0~,\\
&\text{polar voice: }& k \ne 0 \text{ and } n = 0~,\\
&\text{radial voice: }& k = 0 \text{ and } n \ne 0~,\\
&\text{mixed voice: }& k \ne 0 \text{ and } n \ne 0~.
\end{align}
\end{subequations}
For the spectra computed in Ref.~\cite{drasco hughes 2006}, the distribution of power among these voices 
was more strongly dependent on eccentricity than on the other orbital parameters.  From the spectra 
in Fig.~\ref{simple spectra}, one might guess that these sets of frequencies are the best spectral classification 
scheme.  The most significant of the mode families dominating simple spectra is exactly the azimuthal voice, 
and the other two dominant mode families are given by one member of either the radial or polar voices.  
For less simple orbits though, the voices defined above will prove a poor means of classifying spectra.  
For most generic orbit geometries, the bulk of the power is carried by the mixed voice, and there will prove to be 
a simple way of grouping the different members of that very large collection of modes.   

\begin{figure}
\includegraphics[width = .459\textwidth]{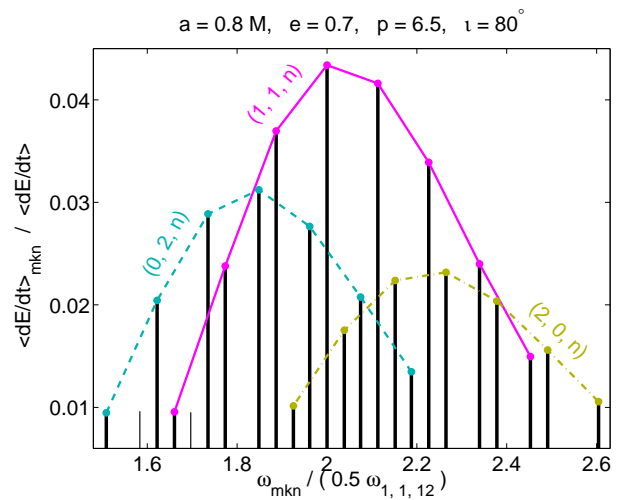}
\caption{\label{hard spectrum} The dominant portion of the spectrum from a highly eccentric, and 
highly inclined, orbital configuration.  The parameters describing the system (black hole spin $a$, 
eccentricity $e$, semilatus rectum $p$, and inclination $\iota$) are again shown at the top of the plot.  
For this spectrum, 4\% of the total power is carried by the dominant mode, 99\% is carried by the top 726 modes, and about 
50\% is carried by the lines shown.  Here the highlighted 
mode families have frequencies $\omega_{11n} = \omega_\phi + \omega_\theta + n\omega_r$ 
(solid magenta line), $\omega_{02n} = 2\omega_\theta + n\omega_r$ (dashed cyan line), 
$\omega_{20n} = 2\omega_\phi + n\omega_r$ (dash-dot mustard-colored line), for various values of $n$.  
All but two of the shown lines are contained in these families (the two excluded lines are drawn slightly 
thinner than the others).}
\end{figure}

The dominant lines of a third sample spectrum (also computed to a fractional accuracy 
of $\varepsilon_\text{flux} = 10^{-6}$, in the total radiated power $\left<dE/dt\right>$) is shown in 
Fig~\ref{hard spectrum}.  
The orbit for this spectrum is both highly eccentric and highly inclined.  The frequencies of 
the dominant modes are also not given by Eq.~(\ref{simple families}), but are instead 
\begin{equation} 
\label{hard families}
\omega = \left\{ 
\begin{array}{ll}
        \omega_\phi + \omega_\theta + n\omega_r & \text{ dominant power at } n=12 \\
        2\omega_\theta + n\omega_r & \text{ dominant power at } n=11 \\
        2\omega_\phi + n\omega_r & \text{ dominant power at } n=14
\end{array}
\right. ~.
\end{equation}
These mode families are not easily classified by the voices (\ref{voices}) of Ref.~\cite{drasco hughes 2006}.
The values $n_{\max}$ of $n$ for the dominant members of these mode families can be crudely approximated 
(to within about 20\% to 40\% for the three families highlighted in Fig.~\ref{hard spectrum}) using the 
conjecture (\ref{PM peaks}).  Equation (\ref{PM peaks}) is a good approximation to the peaks in the spectra derived by Peters and Mathews \cite{peters mathews}
\begin{align}\label{PM power}
\left\langle \frac{dE}{dt} \right\rangle^{\text{PM}}_{\hat n}\!\!
&\propto  \frac{{\hat n}^4}{32} \biggl\{ \bigl[ J_{{\hat n}-2}({\hat n}e) - 2eJ_{{\hat n}-1}({\hat n}e) &
\nonumber \\
& + \frac{2}{{\hat n}}J_{\hat n}({\hat n}e)+2eJ_{{\hat n}+1}({\hat n}e)-J_{{\hat n}+2}({\hat n}e) \bigr]^2&
\nonumber \\
&+ (1-e^2)\left[ J_{{\hat n}-2}({\hat n}e) - 2J_{\hat n}({\hat n}e)+J_{{\hat n}+2}({\hat n}e) \right]^2&
\nonumber \\
& + \frac{4}{3{\hat n}^2}\left[ J_{\hat n}({\hat n}e) \right]^2 \biggr\} ,&
\end{align}
where here $\hat n$ is the multiplier of the single 
frequency of a Newtonian orbit with eccentricity $e$, and $J_{\hat n}(x)$ are Bessel functions of the first kind.
Though a very crude estimator for the values of $n$ describing the dominant members of various mode families, this 
formula is better than any other estimator that has been tried in the present work.  

\begin{figure}
\includegraphics[width = .459\textwidth]{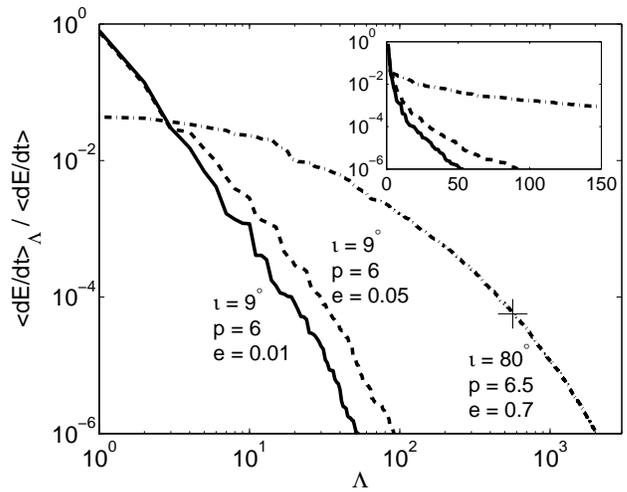}
\caption{\label{falloff} The normalized mode amplitudes for the orbital configurations whose corresponding spectra are shown 
in Figs.~\ref{simple spectra} and \ref{hard spectrum}, sorted in order of decreasing power. The parameters describing the 
orbital geometry (eccentricity $e$, semilatus rectum $p$, and inclination $\iota$) are shown next to each curve.  The crosshairs 
shown on the solid curve indicate the location of the mode with frequency $\omega = 2\omega_\phi$.  That same mode is 
the dominant mode for both of the other systems, as can be read off of Fig.~\ref{simple spectra}.  The inset plot shows the same 
curves, but on a log-linear scale over a shorter range.}
\end{figure}

By comparison to the simple spectra, the number of modes needed to capture any fraction of the total power 
from the complicated spectrum is much larger.  This is demonstrated more clearly by Fig.~\ref{falloff}.
There one can see that, when the modes of a given spectra are sorted by mode number $\Lambda$ in order of decreasing power, mode power falls off roughly as a power law $\Lambda^\alpha$, where $\alpha$ is orbit dependent, for the simple orbits.  For the two simple orbits, the 
rate of this power-law falloff is faster for the less eccentric orbit.  If there is a power-law falloff for the more complicated orbit, it is not 
evident in the first thousand modes.  However, given that the essentially blind algorithm from Ref.~\cite{drasco hughes 2006} computed $\approx$ 30,000 modes before finding the dominant thousand or so shown for the complicated orbit in Fig.~\ref{falloff}, most of the total power is still captured by a surprisingly small number of modes. 

Note also that the mode which dominates the simple orbits, with frequency $2\omega_\phi$, hardly contributes to the spectrum 
of the complicated orbits.   In Fig.~\ref{falloff}, the location of that mode on the curve for the complicated orbit is indicated with a crosshairs at about $\Lambda = 600$, and its power relative to the total power $\approx 6 \times10^{-5}$, is effectively
 insignificant.  This is an extreme example of an effect that has been noticed in simulations of black hole binary inspirals with mass ratios near unity.
In the language of post-Newtonian descriptions of such systems (for a recent overview see Ref.~\cite{van den broeck and sengupta}) all modes other than the one with frequency $2\omega_\phi$ 
are ``higher harmonics.''  The excitation of higher harmonics in those systems has been found to be significant 
in both analytic parameter estimation studies for LISA \cite{sintes vecchio, hellings moore, arun et al, arun et al 2, trias sintes} and fully relativistic numerical simulations \cite{berti et al 2007, vaishnav et al}.  
LISA parameter estimation studies have to date included either spin precession effects \cite{cutler,vecchio,berti buonanno will,lang hughes} or higher harmonics \cite{hellings moore, arun et al, trias sintes}, but not yet both.

\section{Survey of many spectra}\label{Survey of many spectra}

\begin{figure}
\includegraphics[width = .459\textwidth]{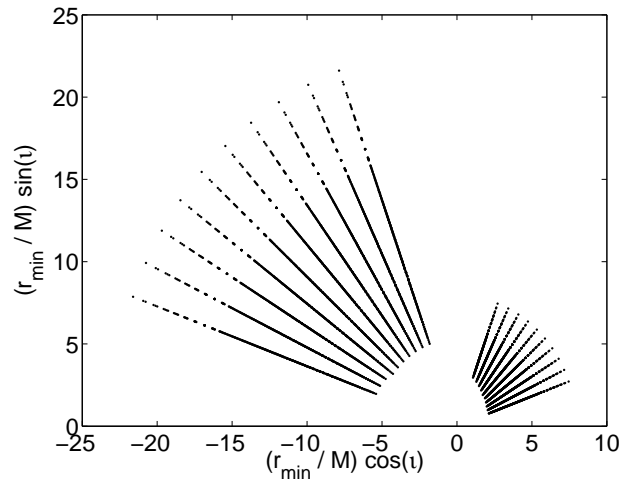}
\caption{A graphical representation of the stable orbits described by the parameters shown in Table 
\ref{OrbitTable}.
The inclination and minimum radius ($r_{\min}$) for the 7,272 stable orbits are shown.}
\label{OrbitPlot}
\end{figure}

In this section, I simulate EMRI snapshots for a large grid of orbital parameters and discuss how the spectral trends 
identified in the previous section vary over the grid.  For each snapshot in the survey, the spin of the 
large black hole is taken to be $a = 0.8M$, and the spectra are computed with the code described in Ref.~\cite{drasco hughes 2006} 
using a requested fractional accuracy of $\varepsilon_\text{flux} = 1\%$, in the total radiated power $\left<dE/dt\right>$.  The grid of 
orbital parameters is uniformly  spaced in eccentricity $e$, 
inclination\footnote{It might be more natural to use a uniform distribution in $\cos \iota$.
However, this would only be a small effect on the actual values of inclination used.  For example, the prograde orbits had ten inclinations uniformly distributed in $\iota$.  If ten equally spaced values of 
$\cos \iota$ were used instead, the average difference in $\iota$ for any of the orbits would have been about $3^\circ$, with the maximum difference being about $6^\circ$.  These changes would not significantly affect any of the conclusions in this work.} 
$\iota$, and in the ratio of the semilatus rectum $p$ to its value for 
the innermost stable circular orbit (ISCO) $p_{\text{ISCO}}$.  The specific values for these parameters are given in Table \ref{OrbitTable}.
\begin{table}
\begin{tabular}{l  l  l | l  l  l} 
$e$    & $p$    & $\iota$        & $e$    & $p$    & $\iota$        \\ \hline
0.1    & 3.4880 & 20$^\circ$     & 0.1    & 10.118 & 110$^\circ$    \\
0.1444 & 3.7632 & 25.556$^\circ$ & 0.1444 & 10.917 & 115.56$^\circ$ \\
0.1889 & 4.0388 & 31.111$^\circ$ & 0.1889 & 11.716 & 121.11$^\circ$ \\
0.2333 & 4.3140 & 36.667$^\circ$ & 0.2333 & 12.514 & 126.67$^\circ$ \\
0.2778 & 4.5893 & 42.222$^\circ$ & 0.2778 & 13.313 & 132.22$^\circ$ \\
0.3222 & 4.8648 & 47.778$^\circ$ & 0.3222 & 14.112 & 137.78$^\circ$ \\
0.3667 & 5.1401 & 53.333$^\circ$ & 0.3667 & 14.911 & 143.33$^\circ$ \\
0.4111 & 5.4157 & 58.889$^\circ$ & 0.4111 & 15.710 & 148.89$^\circ$ \\
0.4556 & 5.6909 & 64.444$^\circ$ & 0.4556 & 16.509 & 154.44$^\circ$ \\
0.5    & 5.9662 & 70$^\circ$     & 0.5    & 17.307 & 160$^\circ$    \\
0.54   & 6.2417 &                & 0.54   & 18.106 &  \\
0.58   & 6.5170 &                & 0.58   & 18.905 &  \\
0.62   & 6.7922 &                & 0.62   & 19.703 &  \\
0.66   & 7.0678 &                & 0.66   & 20.503 &  \\
0.7    & 7.3431 &                & 0.7    & 21.301 &  \\
0.74   & 7.6186 &                & 0.74   & 22.101 &  \\
0.78   & 7.8939 &                & 0.78   & 22.899 &  \\
0.82   & 8.1691 &                & 0.82   & 23.698 &  \\
0.86   & 8.4447 &                & 0.86   & 24.497 &  \\
0.9    & 8.7199 &                & 0.9    & 25.295 &       
\end{tabular}
\caption{\label{OrbitTable} 
The parameters characterizing the 8000 orbit geometries considered (all about a black hole with spin $a = 0.8M$).
The prograde ($\iota < 90^\circ$) are characterized by the 4000 possible combinations of the
first three columns, and the retrograde ($\iota > 90^\circ$) are characterized by the 4000 possible combinations of the
last three columns.  The two different ranges for $p$ correspond to one uniform range for $p / p_\text{ISCO}$.
See Fig.~\ref{OrbitPlot} for a graphical representation of the stable orbits.
Note that, for prograde ($\iota < 90^\circ$) orbits, $p_\text{ISCO} \approx 2.9066$.
Retrograde ($\iota > 90^\circ$) orbits have a less relativistic $p_\text{ISCO} \approx 8.4318$.}
\end{table}
Of the 8000 orbit geometries shown in Table \ref{OrbitTable},  
728 are unstable.  For the unstable orbits, the derivative of the radial potential 
is negative at the prescribed minimum radius
\begin{equation}
\left. \frac{d}{dr} \left(\frac{dr}{d\lambda}\right)^2  \right|_{r_{\min}} < 0 ~,
\end{equation}
where $\lambda$ is Mino's time parameter, related to proper time $\tau$ by
$d\tau = (r^2 + a^2 \cos^2\theta)d\lambda$, and where the radial potential is 
\begin{equation}
\left(\frac{dr}{d\lambda}\right)^2 = \left( E\varpi^2 - a L \right)^2
  - \Delta\left[\mu^2 r^2 + (L - a E)^2 + Q\right]~.
\end{equation}
Here $\varpi^2 = r^2 + a^2$ and $\Delta = r^2 - 2Mr + a^2$.
Snapshot spectra were computed for each of the remaining 7272 stable orbital configurations, 
represented graphically in Fig.~\ref{OrbitPlot}.  The total computational cost of simulating these spectra
was about 2.7 CPU-years on a machine based on a 3.2 GHz Intel Pentium 4 Xeon processor.

\begin{figure}
\includegraphics[width = .459\textwidth]{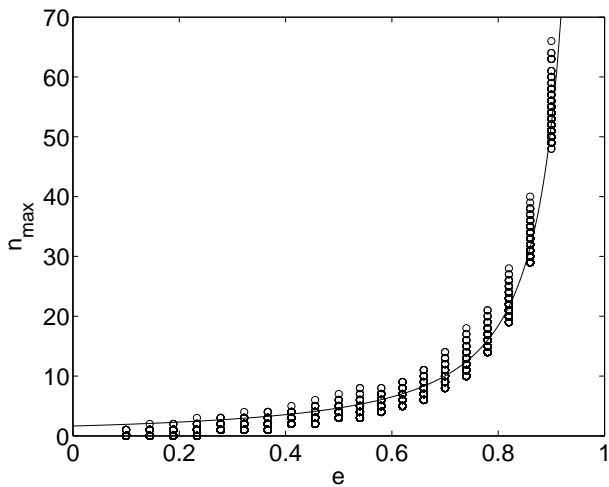}
\caption{The circles indicate the value of $n$ in $\omega_{mkn}$ for the most powerful 
modes in the spectra from the stable orbits in Table \ref{OrbitTable}.    
The solid curve is the approximation (\ref{PM peaks}) that gives the peaks in the 
approximate spectra (\ref{PM power}) derived for Newtonian orbits by Peters and 
Mathews \cite{peters mathews}.}
\label{nmax vs e}
\end{figure}

All of the spectra from this grid are dominated by mode families characterized by 
frequencies
\begin{equation} 
\label{top two families}
\omega = \left\{ 
\begin{array}{l}
	\pm2\omega_\phi + n\omega_r \\
       \pm\omega_\phi + \omega_\theta + n\omega_r  
\end{array}
\right. ~,
\end{equation}
where $\pm$ was 1 for prograde orbits and $-1$ for retrograde orbits.
To within an error on the order of 10\%, the most dominant mode for either family 
had $n = n_{\max}$, where $n_{\max}$ is given by the Peters-Mathews approximation (\ref{PM peaks}).  
The success of that approximation remains similar to that for the more complicated 
sample spectrum in Fig.~\ref{hard spectrum}, and is plotted for the entire set of orbits 
in Fig.~\ref{nmax vs e}.
Of the two dominant mode families (\ref{top two families}) the 
first one, with $(m, k, n) = (2, 0, n)$, is the most common.  The exceptions, dominated by the $(1, 1, n)$-family, 
are the orbits with $\iota$ nearest to 90$^\circ$.  This trend is demonstrated
in Table \ref{elevens}.
\begin{table}
\begin{tabular}{c c} 
$\iota$  & Fraction of orbits  \\ \hline
   20$^\circ$        &  0\\
   25.556$^\circ$ &  0\\
   31.111$^\circ$ &  0\\
   36.667$^\circ$ &  0\\
   42.222$^\circ$ &  0\\
   47.778$^\circ$ &  0\\
   53.333$^\circ$ &  0\\
   58.889$^\circ$ &  26\% \\
   64.444$^\circ$ &  94\% \\
   70$^\circ$        &  100\%\\
  110$^\circ$       &  100\%\\
  115.56$^\circ$  &  100\%\\
  121.11$^\circ$  &  100\%\\
  126.67$^\circ$  &  100\%\\
  132.22$^\circ$  &  1.8\%\\
  137.78$^\circ$  &  0.3\%\\
  143.33$^\circ$  &  0.3\%\\
  148.89$^\circ$  &  0.3\%\\
  154.44$^\circ$  &  0.3\%\\
  160$^\circ$       &  0.3\%
\end{tabular}
\caption{\label{elevens} 
The fraction of orbits, from the grid of orbits described in Table \ref{OrbitTable}, 
dominated by a mode with frequency $\omega_{mkn}$,
where $(m, k, n) = (\pm1, 1, n)$, as a function of inclination $\iota$.
The remaining orbits are dominated by modes with $(m, k, n) = (\pm2, 0, n)$.
Here $\pm$ is $1$ for prograde orbits ($\iota < 90^\circ$) and $-1$ for retrograde orbits ($\iota > 90^\circ$).  Note that since 
the grid of orbits is uniformly spaced in $p/p_\text{ISCO}$, rather than in $p$, the 
prograde orbits have smaller values for $p$ than do the retrograde orbits.}
\end{table}
The prograde orbits tend to be less easily dominated by the $(1, 1, n)$ family.  
Since the orbit grid is evenly spaced in $p/p_{\text{ISCO}}$, and since $p_{\text{ISCO}}$ is much smaller 
for prograde orbits than for retrograde orbits, this suggests that the closer the orbit comes 
to the horizon, the harder it becomes for the system to channel radiation away from the $(2, 0, n)$-family and  
into the $(1, 1, n)$ family.

\begin{figure}
\includegraphics[width = .459\textwidth]{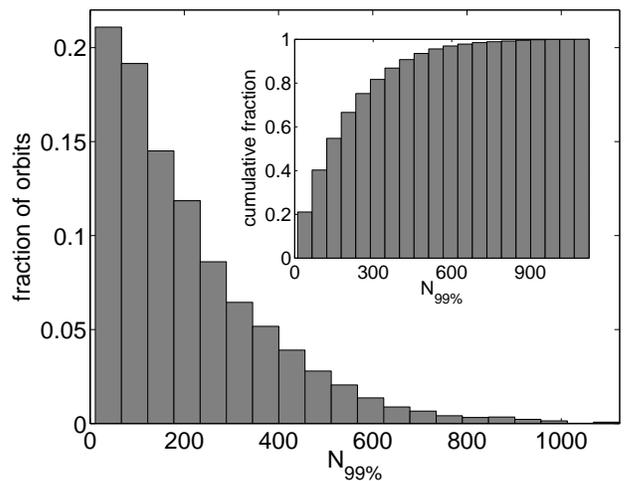}
\caption{A histogram showing the fraction of orbits, out of the 7,272 stable orbits in 
Table \ref{OrbitTable}, for which $N_{99\%}$ modes are required to capture 99\% of the 
radiated power.  The inset displays the same data as a cumulative distribution (its vertical 
axis showing the fraction of orbits with $N_{99\%}$ less than the value on the horizontal axis).}
\label{N99 histogram}
\end{figure}

\begin{figure}
\includegraphics[width = .459\textwidth]{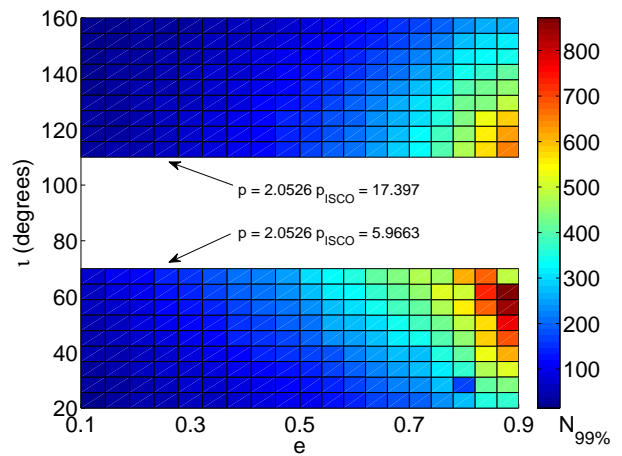}
\caption{The number of modes required to capture 99\% of the 
radiated power, $N_{99\%}$, as a function of eccentricity $e$ and inclination $\iota$
for a fixed value of $p/p_{\text{ISCO}}$.  Since $p_{\text{ISCO}}$ is larger for 
retrograde orbits, the top block of this plot corresponds to orbits with larger values of 
$p$.}
\label{N99 contour}
\end{figure}

As with the sample orbits, the number of modes needed to capture the bulk of the radiated power remains 
resonably small.  This is shown in Fig.~\ref{N99 histogram} by a histogram of the number of modes $N_{99\%}$ carrying 99\% of the power.
An example of the dependence of $N_{99\%}$ on orbit geometry is plotted in Fig.~\ref{N99 contour}.
That plot shows the values of $N_{99\%}$ for all the orbits with the smallest value of $p/p_{\text{ISCO}}$, 
such that over the grid's range for the other orbital parameters, no orbits were unstable.  

It is important to emphasize that the values of $N_{99\%}$ given in this and the following section 
are accurate only to $\sim 10\%$.  This is because $N_{99\%}$ is found only after the algorithm from Ref.~\cite{drasco hughes 2006}
computes a much larger number of modes $N$ in an effort to determine the total power to its requested fractional accuracy 
of $\varepsilon_\text{flux} = 1\%$.  Once that algorithm terminates, the modes that it computed are sorted in order of decreasing 
power such that
\begin{equation}
\left< \frac{dE}{dt} \right> = \sum_{\Lambda = 1}^N \left< \frac{dE}{dt} \right>_\Lambda~,
\end{equation}
where $\left<dE/dt\right>_\Lambda$ decreases with $\Lambda$.  The value of $N_{99\%}$ is the smallest possible 
value satisfying 
\begin{equation}
\sum_{\Lambda = N_{99\%}}^N \left< \frac{dE}{dt} \right>_\Lambda < (0.01) \left< \frac{dE}{dt} \right>~.
\end{equation}
Spot checking its dependence on $\varepsilon_\text{flux}$ for a few sample spectra gave an expected accuracy $\sim 10\%$.

Figure \ref{N99 contour} also shows that, as with the sample orbits, $N_{99\%}$ is more strongly dependent on eccentricity 
than on inclination and semilatus rectum.  This is especially significant for the prospect of observing intermediate mass ratio inspirals with ground-based detectors like LIGO, since the systems that have been estimated to be the most likely candidates for 
being observed are those with especially small eccentricity, typically $e < 10^{-4}$ and at most $e\approx 0.1$ \cite{mandel et al 2007}.  For LIGO, it is likely that waveforms needed for detection need only
contain a few to $\sim 10$ modes.  This both simplifies LIGO's task of detection, since the waveform snapshots will not be very complicated, and hardens its task of spacetime mapping, since correspondingly less information will be observable.

\section{Spectra from a kludged inspiral}\label{Spectra from a kludged inspiral}

In this section, I describe how general spectral characteristics should be expected to evolve during an EMRI.  
The snapshots studied here will be sampled at approximately 12-hour intervals over the final three years of a
single kludged EMRI thought to be typical of the kind that could be observed by LISA.  
The kludged trajectory through orbital parameter space $e(t)$, $\iota(t)$, and $p(t)$, was 
provided by Jonathan Gair, and was numerically computed according to the prescription introduced in 
Ref.~\cite{gair glampedakis 2005}.  
Their method for approximating the trajectory is based on an eclectic combination of approximations 
including post-Newtonian equations
for the radiative fluxes of energy and angular momentum, numerical fits to Teukolsky-based calculations of fluxes for circular 
and equatorial orbits, as well as some uncontrolled approximations for the evolution of Carter's constant.   Since
only power spectra will be discussed here, the results are independent of any evolution for the positional orbital elements, or
$\chi_{mkn}$ in Eq.~(\ref{positional phase}).
While one would expect minimal accuracy from such an array of approximations, these kludged trajectories 
have been shown to exhibit a stunning degree of agreement with more accurate calculations.  
The integrated overlap between approximate waveforms based on the kludged orbit trajectories
and waveforms constructed from black hole perturbation theory alone is often about 95\% \cite{babak et al 2007}.  
So a kludged orbit trajectory is likely more than adequate for the present purpose, since the snapshots 
themselves will still be accurate up to leading order in the mass ratio, and since for the majority of the spectra examined in 
the previous section, the general character of a spectrum does not change dramatically for small perturbations to the orbital 
parameters.  

\begin{figure}
\includegraphics[width = .459\textwidth]{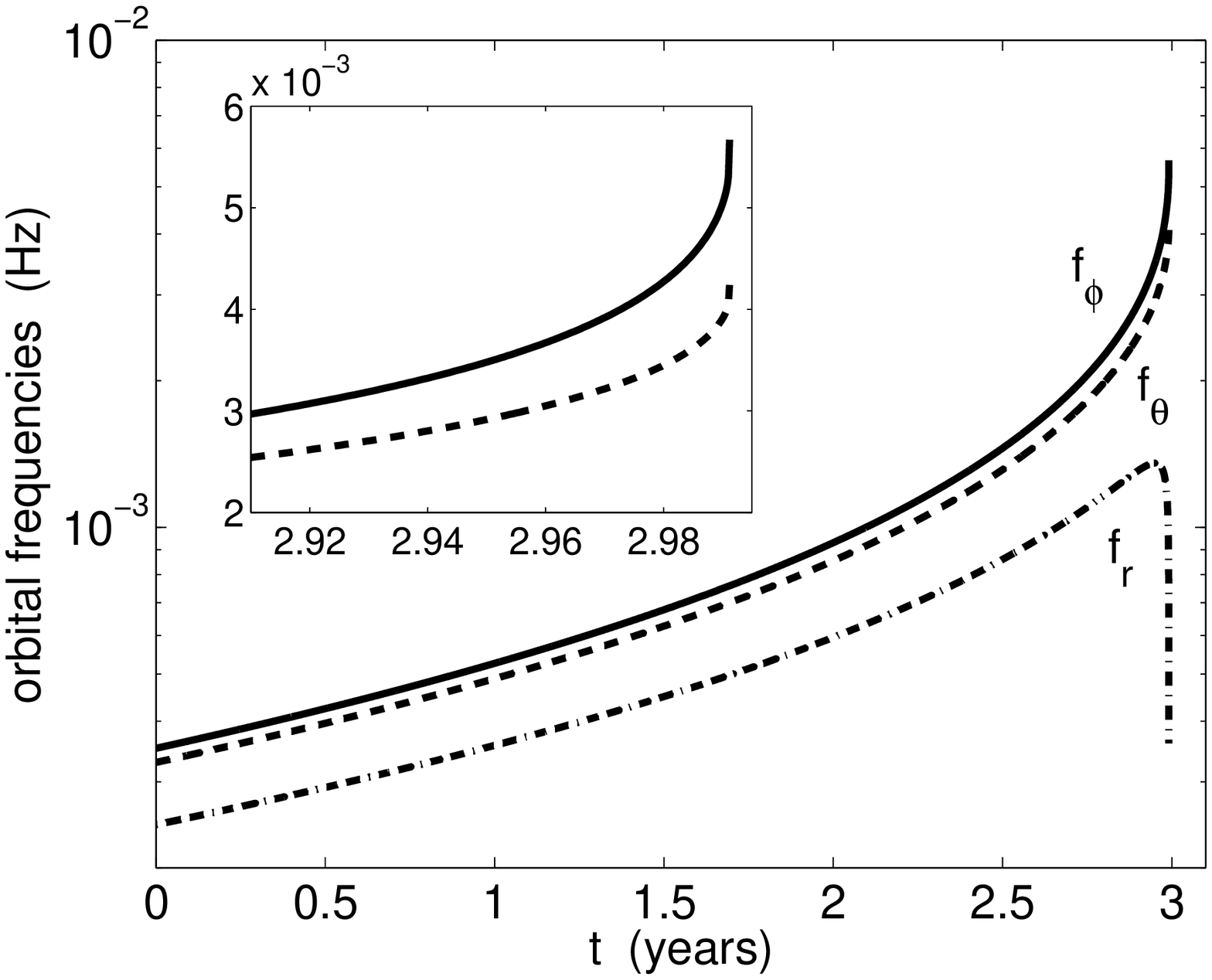}
\includegraphics[width = .459\textwidth]{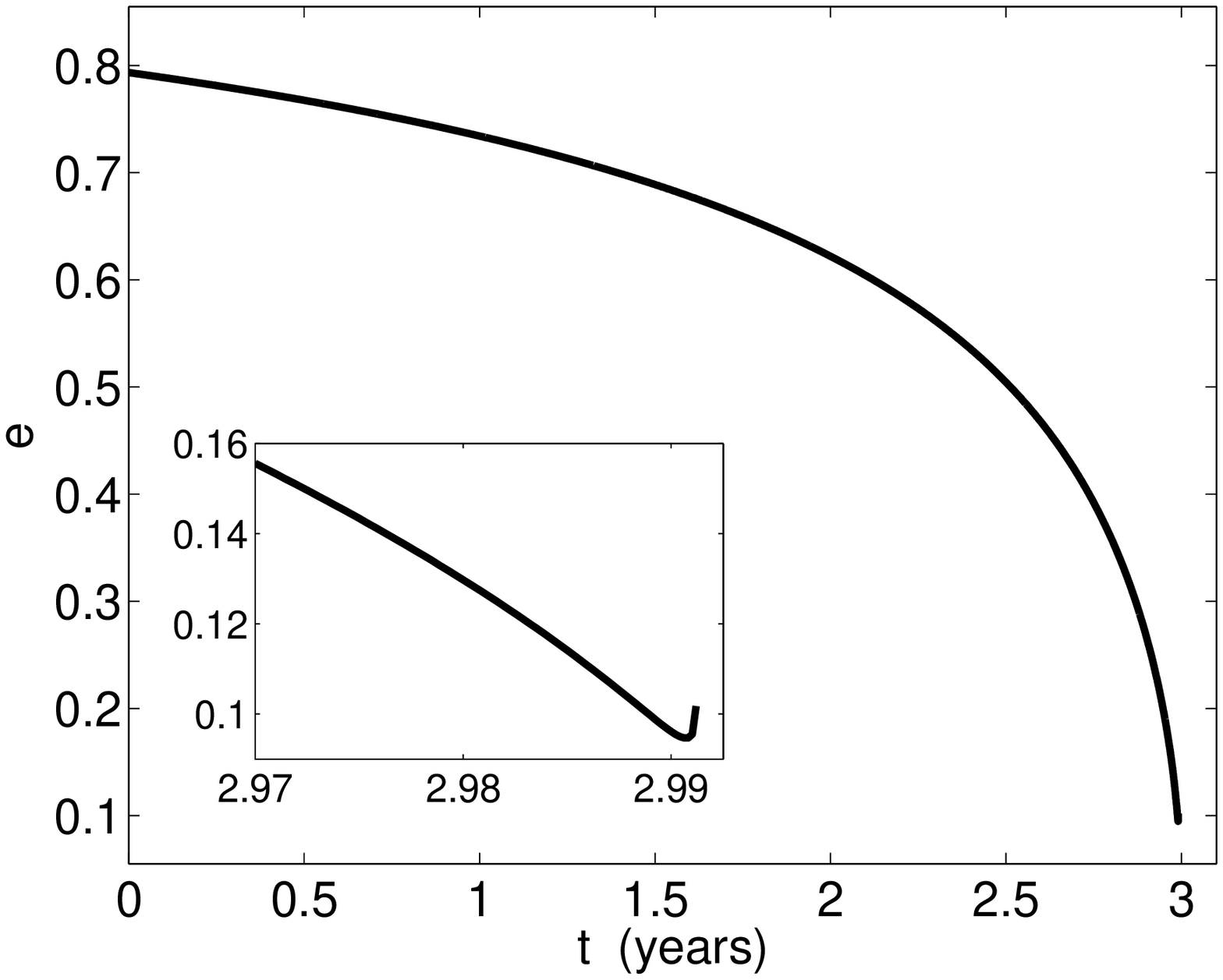}
\caption{\label{kludge} The kludged evolution for a selection of the orbits principal 
elements during an inspiral of the sort that could be observed with LISA.  For this system, the 
large black hole has mass $M = 10^6 M_\odot$, and the magnitude 
of its spin angular momentum is $a = 0.9M$.
The smaller black hole is nonspinning and has mass $\mu = 10 M_\odot$.  
The initial orbit has eccentricity $e = 0.79$, semilatus rectum $p = 8.8$, and inclination $\iota = 43^\circ$.  
The final orbit has eccentricity $e = 0.10$, semilatus rectum $p = 3.2$, and inclination $\iota = 45^\circ$.
The orbit trajectory $e(t)$, $p(t)$, and $\iota(t)$ for this inspiral was computed by Jonathan Gair using the 
algorithm described in Ref.~\cite{gair glampedakis 2005}.  The approximations used to compute 
the trajectory should become less accurate with time as the system becomes increasingly relativistic.
The top panel shows the evolution of the orbital frequencies $f_{r,\theta,\phi} = \omega_{r,\theta,\phi}/(2\pi)$ 
with an inset of the same but over a different range.  The bottom panel shows the evolution of orbital eccentricity $e$, 
with an inset of the same but over a different range.}
\end{figure}

The kludged orbit trajectory for the inspiral examined here is shown in Fig.~\ref{kludge}\footnote{
There is no special reason for using a black hole spin of $a = 0.9M$ in the kludged trajectory 
as opposed to $a = 0.8M$, as is used for the grid of orbits in the previous section.  The 
difference is due to the trajectory and the present work not being produced in parallel.}
The principal orbital elements evolve slowly and smoothly throughout the majority of the inspiral.  
In the final few days of the inspiral, where the trajectory should be least accurate and where the 
adiabatic approximation itself should begin to fail, the eccentricity begins to rise with time, and the orbital 
frequencies rapidly diverge from each other.  The trajectory ends when it has evolved onto an unstable orbit, 
at which point the binary would merge to form a single black hole.

\begin{figure}
\includegraphics[width = .459\textwidth]{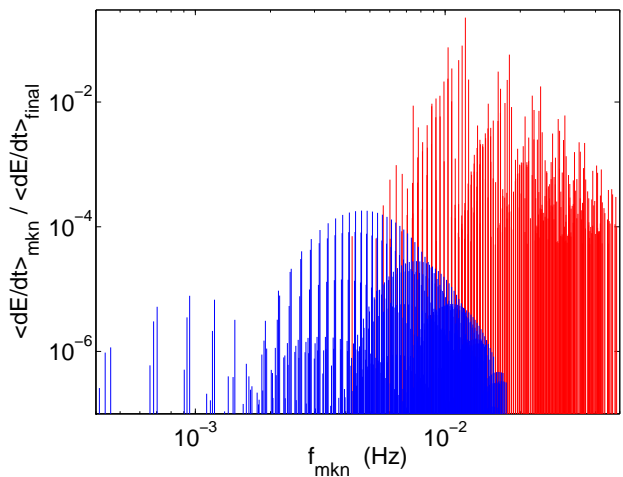}
\caption{\label{first and last spectra} The initial (blue, or dark grey) and final (red, or light grey) power spectra for the inspiral 
shown in Fig.~\ref{kludge}.  An animated movie of the complete spectral evolution is available upon request.}
\end{figure}

Snapshot spectra were computed at roughly 12-hour 
intervals along the orbit trajectory in Fig.~\ref{kludge}.
As was done for the grid of spectra in the previous section, these spectra were computed with the code described in 
Ref.~\cite{drasco hughes 2006} using a requested 
fractional accuracy of $\varepsilon_\text{flux} = 1\%$, in the total power $\left<dE/dt\right>$.  
There were 2,143 snapshots in all, and the total computational 
cost of simulating them was about 1.5 CPU-years on a machine based on a 3.2 GHz Intel Pentium 4 Xeon processor.
The initial and final snapshot spectra from this sequence are shown in Fig.~\ref{first and last spectra}.
The inspiral's initial spectrum is similar in character to the one shown in Fig.~\ref{hard spectrum}, due to the large initial orbital
eccentricity.  The final spectrum is more similar to the ones in Fig.~\ref{simple spectra}. 
The dominant mode families for both the initial and final spectra have $(m, k, n) = (2, 0, n)$ 
and $(1, 1, n)$. Those two families make up the two largest arcs or lobes of lines in the initial spectrum.  
In the final spectrum, the lobes are much more narrow, and are not as easily identified by eye.  For the final spectrum,
a more efficient definition of mode families might instead use $m$ as the free index, rather than $n$.    For example, 
the strongest lines along the upper right edge of the final spectrum are $(m, 0, m)$, peaked at $m = 2$.

\begin{figure}
\includegraphics[width = .459\textwidth]{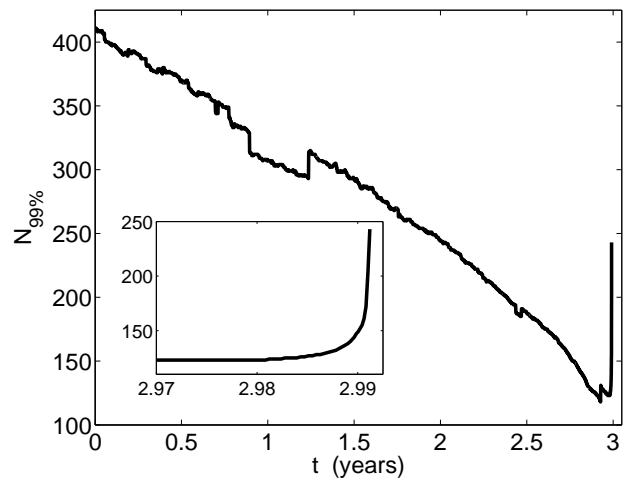}
\caption{\label{KITS N99} The evolution of the number of modes needed to capture 99\% of the 
radiated power $N_{99\%}$ for the inspiral shown in Fig.~\ref{kludge}. The inset shows a close-up view
over about the last week of the inspiral.}
\end{figure}

Figure \ref{KITS N99} shows the evolution in spectral complexity during the inspiral.  
As the binary becomes more circular, the number of lines carrying 99\% of the spectral power 
$N_{99\%}$ decreases with time until the last few days, 
at which point both the kludge and adiabatic approximations should fail.  Over those last few days 
the orbital eccentricity rises and $N_{99\%}$ jumps from about 125 to 250.  
This effect is typical of both kludged trajectories and the more accurate Teukolsky-based 
trajectories. While it may be a physical effect, it always occurs suspiciously in regimes where the adiabatic 
approximation and kludge should be least accurate.  The question of whether or not it is physical  
might best be addressed by the numerical relativity community.  The causes of the smaller 
abrupt jumps in $N_{99\%}$ (for example, at either end of a six month gap centered at about one year) 
is unknown, however, they are within the expected accuracy $\sim 10\%$ whereas the final jump from 125 to 250 is not.  
Repeating the snapshot calculations with a smaller requested overall accuracy in the total power $\varepsilon_{\text{flux}}$ 
would likely eliminate most, if not all, of these small jumps.

\section{Verification of black hole orbits}\label{Verification of black hole orbits}

In this section I suggest EMRI detection algorithms based on the spectral trends found above and 
discuss the scientific meaning of successful detections.  The tone of the discussion is meant to be exploratory 
rather than exhaustive.  That is, it is meant to outline the sorts of data analysis strategies that the 
simulations described above suggest would be useful, rather than to look in great detail at any one algorithm.  

This section will be divided up into two subsections.  The first subsection outlines algorithms that could search for 
systems without radiation reaction.  These are really just of interest as building blocks for more complicated searches
since only a small subset of EMRIs would be observable without accounting for the influence of radiation on the binary.  
The second subsection describes how these building blocks would be used to search for realistic systems affected by 
radiation reaction and gives a detector-independent estimate of how well these algorithms might 
perform in the best of circumstances.

\subsection{Without radiation reaction}

For times that are sufficiently short for the orbit of the captured mass to be unaffected by radiation\footnote{
For a subset of observable systems ~\cite{drasco 2006} these ``short'' times are actually longer than the longest amounts of time for which 
fully coherent search algorithms are computationally affordable (a few weeks \cite{gair et al 2004}).
}, the general expression for EMRI snapshots (\ref{waveform model}) is accurate up to corrections that are at most 
of order $\mu^2/M^2$ and are perhaps even as small as $\sim \mu^3/M^3$ with the principal frequencies adjusted according 
to the second order metric perturbation \cite{mino 2007}.  In this subsection, I will outline a detection strategy for a 
signal of this form.  Truncating the general expression for an EMRI-snapshot waveform (\ref{waveform model}) to a finite number 
of $N$ modes, rewriting it as a single sum, and explicitly showing the positional phase elements, gives the following
\begin{eqnarray}\label{truncated model}
h_{+} - i h_{\times} &=& 
	\sum_{\Lambda=1}^N h_{\Lambda} \exp\left\{-i [m_\Lambda \omega_\phi (t-t_\phi) \right. \nonumber \\
	&+&\left. k_\Lambda \omega_\theta (t-t_\theta)+ n_\Lambda \omega_r (t-t_r)]\right\}~.
\end{eqnarray}
I wish to consider the prospects of using this truncated expression (\ref{truncated model}) explicitly as a
phenomenological template.  

By phenomenological templates, I mean templates for which some of the 
free and measurable parameters will have no immediately obvious physical meaning.  Traditional, or nonphenomenological 
templates are constructed as follows.  You start by declaring the 17 parameters which completely determine 
the waveform (e.g.~the two masses, and 3-vectors for the position and velocity of each object as well as for the 
spin of the larger object).  You then solve Einstein's equation or some perturbed form of it to 
completely determine $h_{+,\times}$.  This last step is equivalent to determining all of the frequencies $\omega_{r, \theta, \phi}$, 
phase shifts $t_{r, \theta, \phi}$, and Fourier coefficients $h_{\Lambda}$ in the model (\ref{waveform model}).  
The phenomenological templates are computed in a much simpler way.  You first chose values for the frequencies, phase shifts, and 
Fourier coefficients, and then you simply evaluate the sum over modes (\ref{truncated model}).  
For these waveforms, the frequencies, phase shifts, and Fourier coefficients are themselves the free quantities 
to be measured.  For a phenomenological template with $N$ modes, there are then $5N+6$ free parameters.  Of those there are 
$2N+6$ real numbers (the three frequencies, the three phase shifts, and the $N$ complex mode amplitudes), as well as $3N$ 
integers (the frequency multipliers $m_\Lambda$, $k_\Lambda$, and $n_\Lambda$).

The most obvious objection to the use of these phenomenological templates in gravitational wave searches 
is the large dimensionality of the parameter space.  
For templates that allow for any more than $N=2$ modes, you will have more free parameters than the 
traditional templates\footnote{The specific value $N=2$ here is not exact.  The three integer parameters are 
much simpler degrees of freedom from the standpoint of a search (discrete and reasonably confined).  
In the same vein though, not all of the 17 traditional parameters are intrinsic.  So while the value of $N$ 
for which both the traditional and phenomenological template spaces are dimensionally equivalent is surely small, 
the exact value is not obvious.}, and of those only the three 
frequencies $\omega_{\phi,\theta,r}$ will carry immediate physical meaning.  
In the traditional scheme, however, the step of going from the complete set of 17 parameters to the waveform is 
extremely costly.  Since both template families are equally good matches to the true waveforms, and since either way you will be dealing 
with a significant number of template parameters, it may be worth adding dimensionality to the template parameter space in exchange 
for not having to solve Einstein's equation.  

Though the large dimensionality may seem daunting, especially for those familiar with efforts focused on
sources with just a few degrees of freedom, it is not unrealistic for elaborate algorithms to identify signals characterized by 
a large number of parameters.  For algorithms that match against templates over a sequence of increasingly dense grids on the model 
parameter space, computational cost grows as a power law where the power is proportional to the number of free parameters.  Algorithms 
designed for more complex models have costs that instead grow only linearly with the number of parameters.  Algorithms of this nature have 
been used in the mock LISA data challenges \cite{mldc} to recover $\sim 10^4$ white dwarf binaries.   

I now discuss how well phenomenological detection algorithms can be expected to perform in the best case scenario 
where the algorithm has no difficulty in selecting the correct parameter values.  This can be done without reference 
to specific instruments, and in a sky averaged sense, by studying the distribution of the power among various 
modes relative to the total power radiated by the binary.  More sophisticated estimates that account for detector characteristics 
and variation of signal parameters are certainly possible, but they are beyond the scope of this paper.

For the task of searching only for EMRI snapshots with phenomenological templates, the example spectra from Sec.~\ref{Sample spectra} 
suggest estimates of the minimum scale of the phase space dimensionality.  For example, systems with eccentricities of 
about 1\% will produce spectra similar to Fig.~\ref{simple spectra}.  So a ground-based phenomenological search for these systems would 
require $N\lesssim 10$ modes, resulting in a model with 56 parameters (26 real numbers and 30 relatively small integers).  
For snapshots thought to be typical of EMRIs that could be observed with space-based detectors, one would need $N\sim10^2$ to $10^3$.   
This would mean a model with about 500 to 5000 free parameters,  still far fewer than what has been demonstrated already for white 
dwarf binaries \cite{mldc}.  

If a search for phenomenological EMRI snapshots were successful, only the general waveform model would be verified.  
One would only be able to claim detection of some signal with a discrete triperiodic 
spectrum with some measured fundamental frequencies $\omega_{\phi, \theta, r}$, since the physical meaning of the other unknown 
parameters is convoluted. In this event, one could then turn to the underlying physical model.  Finding a set of its 17 
free parameters that best reproduce the parameters measured in the phenomenological search would then confirm 
the physical model to some more explicit level of uncertainty.  Failing to find parameters for the underlying physical 
model might mean that the snapshot was produced by a test mass moving along a geodesic of some non-Kerr spacetime, since 
many (but not all) candidates for such orbits are also triperiodic \cite{gair li mandel 2007, flanagan hinderer 2007}. 

Some intermediate level of model verification is also possible and could reduce the dimensionality of the parameter space for the
phenomenological templates.  For example, sixteen mode families are needed to capture 
99\% of the power radiated by the initial snapshot from the inspiral examined in the previous section. 
The distribution of power among these families is shown in Table \ref{first families}.
\begin{table}
\begin{tabular}{c c c c} 
$m$  & $k$ & $n_{\max}$ & $ \left<\dot E\right>_{mk} / \left<\dot E\right>$ \\ \hline
   2 &  0 &  16 &  5.1$\times 10^{-1}$ \\
   1 &  1 &  15 &  2.2$\times 10^{-1}$ \\
   3 &  0 &  27 &  9.2$\times 10^{-2}$ \\
   2 &  1 &  25 &  6.2$\times 10^{-2}$ \\
   0 &  2 &  13 &  3.5$\times 10^{-2}$ \\
   4 &  0 &  38 &  2.0$\times 10^{-2}$ \\
   1 &  2 &  24 &  1.7$\times 10^{-2}$ \\
   3 &  1 &  36 &  1.8$\times 10^{-2}$ \\
   2 &  2 &  35 &  6.4$\times 10^{-3}$ \\
   2 & -1 &  12 &  3.7$\times 10^{-3}$ \\
   4 &  1 &  47 &  2.9$\times 10^{-3}$ \\
   5 &  0 &  49 &  2.5$\times 10^{-3}$ \\
  -1 &  3 &  12 &  1.9$\times 10^{-3}$ \\
   0 &  3 &  22 &  1.3$\times 10^{-3}$ \\
   3 &  2 &  45 &  1.5$\times 10^{-3}$ \\
   3 & -1 &  21 &  9.9$\times 10^{-4}$
\end{tabular}
\caption{\label{first families} 
The 16 mode families needed to capture 99\% of the power radiated 
during the first 12 hours of the inspiral discussed in Sec.~\ref{Spectra from a kludged inspiral}.  
Each family is defined by fixing $m$ and $k$.  The most powerful member of each family has 
frequency $m \omega_\phi + k \omega_\theta + n_{\max} \omega_r$.  The ratio of the power 
radiated by each family to the total power from this first 12 hours is given 
by $\left<\dot E\right>_{mk} / \left<\dot E\right>$.  For this system, the spin of the black hole is $a = 0.9M$, 
and the orbit of the test mass has eccentricity $e = 0.79$, semilatus rectum $p = 8.8$, and inclination $\iota = 43^\circ$.}
\end{table}
As is true for all the snapshots simulated in this paper, this one is dominated by the mode families with $(m, k, n) = (2, 0, n)$ and 
$(1, 1, n)$.  And as is typical, those two mode families carry most of the power, 73\% here.  In an effort to simplify the 
phenomenological waveform model, one might restrict it to include only those mode families.  
This specific model would eliminate ($m_{\Lambda}$, $k_{\Lambda}$) from the template parameter space by fixing them 
to either $(2, 0)$ or $(1, 1)$, and would create two new parameters specifying the number of modes in each of the two families.
This would reduce the number of free parameters from $5N+6$ to $3N+8$.

\subsection{With radiation reaction}

The scenario describe in the previous subsection is only immediately useful for EMRI's with the most extreme
mass ratios.  There is no compelling reason to expect those systems to be especially common, 
or to even consider them reasonable targets at all.  The purpose of studying these simple systems is to 
construct from the results an approximate description of more generic EMRIs that respond to their own radiation.  
That is, we wish to describe the radiation of a generic adiabatic EMRI as a slowly evolving sequence of EMRI 
snapshots.  Here I will now outline how the phenomenological templates for EMRI snapshots could be modified to 
describe the more general class of adiabatic EMRIs.  There are many ways that this could be done.  
Although a detailed study of specific models would be valuable, it is beyond the scope of this paper.  
Instead, I aim to be as general as possible and will steer away from discussing any 
specific implementation.

For an adiabatic EMRI, the motion of the small object is described by a solution of the geodesic equation for 
the Kerr spacetime, but with the orbital elements replaced by quantitates that evolve slowly.  The Teukolsky equation
can provide the leading order radiative changes to those quantities, and more sophisticated techniques are envisioned
for describing both conservative and radiative effects.  In the spirit of trading calculation difficulty for added dimensionality, every quantity that was constant for the snapshot model (\ref{truncated model}) could in principle be
replaced by simple, one or two parameter models. 

To illustrate this, consider the orbital frequencies $\omega_i$, for $i = r$, $\theta$, $\phi$.
These can be taken to drift linearly with time
\begin{equation}
\omega_i \to \omega_{i} + \dot \omega_{i} t~.
\end{equation}
where I have introduced new constants $\dot \omega_{i}\propto \mu/M$.  
This simple model fits the first year of the three frequency 
trajectories shown in Fig.~\ref{kludge} with an average fractional accuracy of about 1\%.  
Other models can of course do better.  For example, Peters and Mathews derived an expression for radius as a function 
of time in the case of slow circular inspirals.  Combining that result, 
Eq.~(5.9) of Ref.~\cite{peters mathews}, with Kepler's law $M\omega = (r/M)^{-2/3}$, gives 
\begin{equation}
M\omega = \left[\left(\frac{r_0}{M}\right)^4 - \frac{256}{5}\frac{\mu}{M}\frac{t}{M} \right]^{-3/8}~.
\end{equation}
For the more general case of fast generic motion, one might want to try a model with a similar form
\begin{equation}
M\omega_i = [\alpha_i + \beta_i (t/M)]^{-\gamma_i}~.
\end{equation}
This model fits the first year of the three frequency trajectories shown in Fig.~\ref{kludge} with 
$\gamma_i \approx 1.6$ and an average fractional accuracy on the order of $10^{-4}$.
It performs similarly at later times, but not with the same values of the parameters 
$\alpha_i$, $\beta_i$, and $\gamma_i$.  Unlike the simple linear model however, this one has 
three parameters instead of two. 

To complete the construction of phenomenological templates for adiabatic EMRIs, 
similar models can be concocted for the other parameters of the snapshot 
templates (\ref{truncated model}), the complex mode amplitudes, and the positional phase elements.  
If one-parameter models like the linear frequency drift are used, then the new template is given 
by Eq.~(\ref{truncated model}) with
\begin{eqnarray}
h_\Lambda &\to& h_\Lambda + \dot h_\Lambda t ~,\\
m_\Lambda &\to& m_\Lambda + \dot m_\Lambda t ~,\\
k_\Lambda &\to& k_\Lambda + \dot k_\Lambda t ~,\\
n_\Lambda &\to& n_\Lambda + \dot n_\Lambda t ~,\\
\omega_i &\to& \omega_{i} + \dot \omega_{i} t~. \\
t_i &\to& t_i + \dot t_i t~,
\end{eqnarray}
again for $i = r$, $\theta$, $\phi$.  For this case, the dimensionality of the template parameter space is 
doubled to $10 N + 12$.  Of those, only the $3N$ frequency multipliers are integers.
These phenomenological templates are similar in nature to the time-frequency search methods which have 
been successfully demonstrated in the mock LISA data challenge \cite{gair et al 2007}.  They differ 
from those in that they can accommodate coherent integration and could also more naturally include 
specific schemes for evolving both the principle and positional orbital elements.

\begin{figure}
\includegraphics[width = .459\textwidth]{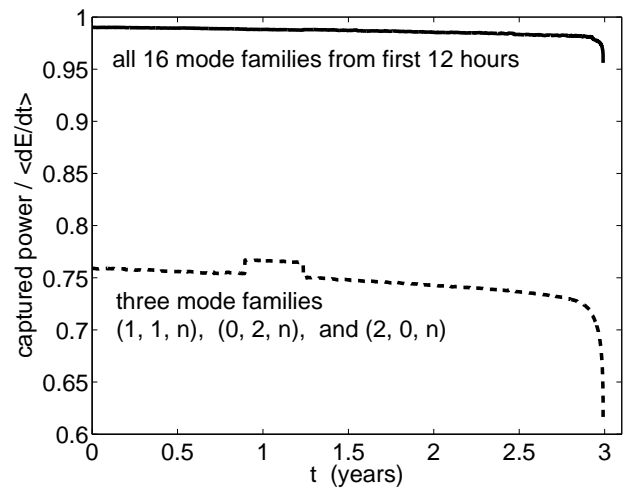}
\caption{\label{captured power} A time-dependent measure of success for two hypothetical detection 
algorithms searching for the inspiral from Fig.~\ref{kludge}.  The algorithm for the dashed blue 
curve captures the power radiated at all frequencies of the form $\omega_{mkn}$, where 
$(m, k, n) = (1, 1, n)$, $(0, 2, n)$, and $(2, 0, n)$, for any $n$. The solid curve captures power 
radiated at all frequencies in the 16 mode families needed to capture 99\% of the initial power, 
where here a mode family is defined as all modes with frequencies $\omega_{mkn}$ for some fixed  $m$ and $k$.}
\end{figure}

Given that EMRI snapshots tend only to be dominated by a small number of mode families, it is likely that 
a slightly simpler waveform model could be used for adiabatic EMRIs.  
Figure \ref{captured power} demonstrates how successful two such hypothetical detection algorithms might be
if searching for the kludged inspiral from Sec.~\ref{Spectra from a kludged inspiral}.
If one were to search for this waveform using a model that included only the dominant three mode families 
from the sample complicated snapshot in Fig.~\ref{hard spectrum} one would recover $71\%$ of the inspiral's total power.
Assuming this could be done with one-parameter models for the adiabatic evolution of the waveforms parameters, such a model 
would have $6N+18$ free parameters (by eliminating $2N$ frequency multipliers and their $2N$ linear drifts, and by adding 
three new integers specifying how many modes are in each family, as well as their three linear drifts).  
For this example EMRI, this scheme included $N=121$ modes.  So 714   free parameters would have been needed to recover 71\% of the power.  
A similar hypothetical algorithm which captured all the power carried by the 16 
mode families that make up most of that inspiral's initial spectrum (Table \ref{first families}) would recover 98\% of the total 
power with $6N+44$ free parameters (by eliminating $2N$ frequency multipliers and their $2N$ linear drifts, and by adding 
16 new integers specifying how many modes are in each family, as well as their 16 linear drifts).  
For this example EMRI, this scheme included $N=411$ modes.  So 2,510 free parameters would have been needed to recover 98\% of the power.  

It should be emphasized that any gravitational wave detections following from the use of these phenomenological waveform models would not necessarily yield either the parameters that completely determine an EMRI (position, masses, etc) or the spacetime 
map that general relativity predicts is encoded in the radiation.  They would however verify the detection of waveforms predicted to 
be produced when a test mass perturbs a rotating black hole by moving through an adiabatic sequence of its bound geodesic orbits.  
They would also measure the evolution of the three fundamental frequencies throughout that sequence, or equivalently the evolution 
for any other set of principle orbital elements.  It is possible that restricting phenomenological EMRI waveforms to include only 
the mode families that are most commonly dominant in the snapshots simulated here may alone be enough of a constraint to keep EMRIs 
into non-Kerr black hole candidates from triggering a detection.  However, without more work with the snapshots from such EMRIs 
one could not say so with any certainty.  One would have to be content only to have verified that radiation from an adiabatic sequence 
of black hole orbits could have triggered the detection.  

\section{Conclusion}\label{Conclusion}

The number of significant modes in generic EMRI-snapshot spectra has been shown generally to be
much more manageable than one might have guessed from earlier truncation 
algorithms \cite{drasco hughes 2006}.  
This should lead to improved truncation algorithms, which will reduce the cost of future data analysis efforts. 
Such improvements should exploit the trends observed here in the relationship between orbit geometry and spectral signature.  The ability to predict the multiplier $n \approx n_{\max}$ of the radial frequency for the dominant modes 
using a formula based on such simple approximations \cite{peters mathews} is encouraging.  It suggests that many
of the trends in these spectra might be understood analytically using more recent 
tools \cite{moreno-garrido mediavilla buitrago, poisson, ganz et al}.

The detection algorithms that are suggested here for verifying minimal aspects of relativity and black hole 
physics may ultimately be used in future gravitational wave detections.  However, more work is needed to determine
whether or not they are cost-efficient and science-efficient alternatives to traditional search techniques.
Another possibly interesting area for future work is to explore the dependence of the snapshot spectra on the spin of
the larger black hole.  It has been implicitly assumed that the large values of spin considered are somehow representative 
of observable EMRIs.  This is reasonable since the few existing measurements due to modeling x-ray spectra from 
galactic nuclei suggest near-maximal spins \cite{miller, reis et al}.
Still, future work that tests the generality of these parameter values by simulating other EMRIs would be worthwhile.

\begin{acknowledgments}
I am especially grateful to Scott Hughes for providing significant portions of the numerical code used in this work, and to Jonathan Gair for providing the kludged inspiral trajectory. 
I would like to thank Chao Li, Geoffrey Lovelace, and Kip Thorne for discussions that triggered this work.  I also thank Emanuele Berti, Yasushi Mino, and Michele Vallisneri for encouragement and helpful discussions.   
The supercomputers used in this investigation were provided by funding from the JPL Office of the Chief Information Officer.
This research was carried out at the Jet Propulsion Laboratory, California Institute of Technology, 
under a contract with the National Aeronautics and Space Administration and funded through the 
internal Human Resources Development Fund initiative, and the LISA Mission Science Office.
\end{acknowledgments}

\appendix
\section{Bi periodic form of time-dependent spatial coordinates}\label{append}

Here I derive the bi-periodic form (\ref{coordinate orbits}) of the 
spatial Boyer-Lindquest coordinates $r$, $\theta$, and $\phi$ for bound 
geodesics as a function of the time coordinate $t$.

The bi-periodic forms of both the radial and polar coordinates follows immediately from
Sec.~IV of Ref.~\cite{drasco hughes 2004} by replacing $f[r(t),\theta(t)]$ with $r(t)$ 
and $\theta(t)$.  The resulting relationships between the coefficients in the Mino-time series 
expansion (\ref{mino orbits}) and the coordinate-time expansions (\ref{coordinate orbits}) are given by
\begin{eqnarray}
{\tilde r_{kn}} &=& 
	\frac{\Upsilon_r \Upsilon_\theta}{(2\pi)^2\Gamma}
	\int_0^{2\pi/\Upsilon_r} d\lambda_r
	\int_0^{2\pi/\Upsilon_\theta} d\lambda_\theta \nonumber \\
	&\times& 
	\frac{dt}{d\lambda}[r(\lambda_r),\theta(\lambda_\theta)] 
	e^{i(k\omega_\theta + n\omega_r)\delta t(\lambda_r,\lambda_\theta)}\nonumber \\
	&\times& 
	r(\lambda_r)
	e^{ik\Upsilon_\theta\lambda_\theta + in\Upsilon_r \lambda_r}~,\\
{\tilde \theta_{kn}} &=& 
	\frac{\Upsilon_r \Upsilon_\theta}{(2\pi)^2\Gamma}
	\int_0^{2\pi/\Upsilon_r} d\lambda_r
	\int_0^{2\pi/\Upsilon_\theta} d\lambda_\theta \nonumber \\
	&\times& 
	\frac{dt}{d\lambda}[r(\lambda_r),\theta(\lambda_\theta)] 
	e^{i(k\omega_\theta + n\omega_r)\delta t(\lambda_r,\lambda_\theta)}\nonumber \\
	&\times& 
	\theta(\lambda_\theta)
	e^{ik\Upsilon_\theta\lambda_\theta + in\Upsilon_r \lambda_r}~,
\end{eqnarray}
where $r(\lambda_r)$ and $\theta(\lambda_\theta)$ are given 
by their Mino-time series expansions with $\lambda = \lambda_r$ and 
$\lambda = \lambda_\theta$, respectively, 
\begin{align}
&r(\lambda_r) = \sum_n r_n e^{-i \Upsilon \lambda_r}~,&
&\theta(\lambda_\theta) = \sum_k \theta_k e^{-i \Upsilon \lambda_\theta}~,&
\end{align}
and where 
\begin{equation}
\delta t(\lambda_r,\lambda_\theta) = 
\sum_{kn} t_{kn} e^{-ik\Upsilon_\theta \lambda_\theta - in\Upsilon_r \lambda_r}~.
\end{equation}
The $\lambda$-derivative of $t$ can be evaluated analytically as given by 
Carter's first order geodesic equations, or it can be found from
\begin{align}
\frac{dt}{d\lambda}[r(\lambda_r),\theta(\lambda_\theta)] &= \Gamma& \nonumber \\
&- i \sum_{kn}
(k \Upsilon_\theta + n \Upsilon_r) t_{kn}
e^{-ik\Upsilon_\theta \lambda_\theta - in\Upsilon_r \lambda_r}~.&
\end{align}

The derivation of the bi periodic form for $\phi(t)$ is slightly different.  First,
write $t(\lambda)$ as
\begin{equation}
t = \Gamma \lambda + \delta t~.
\end{equation}
Multiplying by $\omega_\phi$, and rearranging terms gives 
\begin{equation} \label{upl}
\Upsilon_\phi \lambda = \Omega_\phi t - \Omega_\phi \delta t~,
\end{equation}
since $\omega_\phi \Gamma = \Upsilon_\phi$.  Now write $\phi(\lambda)$ as 
\begin{equation}
\phi = \Upsilon_\phi \lambda + \delta \phi~,
\end{equation}
and insert the above expression (\ref{upl}) for the first term to find
\begin{align}
\phi = \omega_\phi t + \widetilde{\delta\phi}~,
\end{align}
where 
\begin{equation}
\widetilde{\delta\phi} = \delta \phi - \omega_\phi \delta t~.
\end{equation}
Treating $\widetilde{\delta\phi}$ as a function of $r$ and $\theta$ now
allows it to be used in place of $f[r(\lambda), \theta(\lambda)]$ in 
Sec.~IV of \cite{drasco hughes 2004}.  This gives the following
expression for the coefficients of the bi periodic form of $\phi(t)$:
\begin{eqnarray}
{\tilde \phi_{kn}} &=& 
	\frac{\Upsilon_r \Upsilon_\theta}{\Gamma (2\pi)^2}
	\int_0^{2\pi/\Upsilon_r} d\lambda_r
	\int_0^{2\pi/\Upsilon_\theta} d\lambda_\theta \nonumber \\
	&\times& 
	\frac{dt}{d\lambda}[r(\lambda_r), \theta(\lambda_\theta)] 
	e^{i(k\omega_\theta + n\omega_r)\delta t(\lambda_r,\lambda_\theta)}\nonumber \\
	&\times& 
	\widetilde{\delta\phi}(\lambda_r,\lambda_\theta)
	e^{ik\Upsilon_\theta\lambda_\theta + in\Upsilon_r \lambda_r}~,
\end{eqnarray}
where
\begin{equation}
\widetilde{\delta\phi} (\lambda_r,\lambda_\theta) =
	\sum_{kn}(\phi_kn - \omega_\phi t_{kn}) 
	e^{-ik\Upsilon_\theta \lambda_\theta - in\Upsilon_r \lambda_r}~.
\end{equation}


\end{document}